\documentclass[final,3p,times,authoryear]{elsarticle}

\usepackage{amssymb, amsmath, amsthm, lineno, epsfig, bm, latexsym}
\usepackage[bookmarks,colorlinks=true,linkcolor=blue,citecolor=blue,filecolor=red]{hyperref}
\usepackage{pdflscape}

\journal{Advances in Data Analysis and Classification}

\newtheorem{remark}{Remark}

\newtheorem{lemma}{Lemma}

\usepackage[dvipsnames]{xcolor}
\definecolor{mygray}{gray}{0.98}
\usepackage[framemethod=TikZ]{mdframed}
\usepackage{lipsum}
\mdfdefinestyle{MyFrame}{%
	linecolor=black,
	outerlinewidth=0.005cm,
	roundcorner=18pt,
	innertopmargin=\baselineskip,
	innerbottommargin=\baselineskip,
	backgroundcolor=mygray}

\begin{document}
	\begin{frontmatter}
		\title{Mixture of Censored Partially Linear Experts Models for Symmetric or Asymmetric Data with or without Outlying Observations:\\A novel approach with two-piece scale-mixture of normal distributions}
		\author{Mohsen Maleki\fnref{label1}\corref{cor1}}
        \ead{m.maleki.stat@gmail.com}

        \author{Geoffrey J. McLachlan\fnref{label2}}
        %

        \cortext[cor1]{Corresponding author\fnref{cor1}}
        \address[label1]{Department of Statistics, Faculty of Science, University of Fasa, Fasa, Iran}
        \address[label2]{School of Mathematics and Physics, University of Queensland, Queensland, Australia}
		\begin{abstract}
			Finite mixture of partially linear model and finite mixture of experts models offer flexible tools for analyzing heterogeneous data, especially when covariates influence subgroup structure. Partially linear models further enhance this flexibility by combining linear and nonlinear effects. In many applications, however, censoring and departures from normality complicate estimation. To address these issues, we introduce a semiparametric mixture of censored partially linear expert models based on the two-piece scale mixture of normal family, which can accommodate skewness, heavy tails, and censoring. The proposed framework enables closed-form expressions in the E-step of an ECME algorithm, avoiding MCMC-based approximations and yielding stable maximum likelihood estimates. Simulation studies and real data analyses demonstrate the effectiveness of the proposed method in settings with both heterogeneity and censoring.
		\end{abstract}
	\end{frontmatter}
	
	\section{Introduction} \label{Intro}
    The concept of finite mixture of regressions (FMR), first put forward by \cite{Quandt}, offers a way to model data that arise from multiple hidden subgroups. In this approach, each latent group is assumed to follow its own regression relationship, allowing the overall model to capture unobserved structural differences in the population. The literature on FMR has grown considerably, with substantial methodological developments reported by \cite{Jiang, Mclachlan, Neykov, Garcia, Bai, Bashir, Hunter, Yao, Song, Ma, Zarei, Oh}.

    A well-recognized drawback of traditional FMR models is their reliance on the assumption that covariates play no role in determining cluster membership \citep{Hening}. When this independence assumption is not appropriate, the resulting regression–based clustering may be unreliable. To overcome this issue, \cite{Jacobs} introduced the mixture of linear experts (MoE) framework, in which the mixing proportions are allowed to vary as functions of the covariates. Since then, the MoE model has undergone numerous extensions. \cite{Nguyen} explored the use of Laplace errors, while \cite{Chamroukhi1,Chamroukhi2} examined Student-\textit{t} and skew-\textit{t} error structures. \cite{Murphy} proposed a more parsimonious MoE architecture aimed at improving estimation efficiency. \cite{Mirfarah} incorporated scale mixtures of normal to model the error distribution. More recently, \cite{Oh} introduced a tailored MoE formulation, adding to this expanding line of research.

    For data arising from a single homogeneous population, \cite{Engle} proposed the partial linear model, which blends a parametric linear component with a nonparametric function. In this framework, the response variable $Y$ is modeled as:

    \begin{equation}
    	Y = \boldsymbol{x}^\top \boldsymbol{\beta} + g(t) + \varepsilon,
    	\label{eq:1}
    \end{equation}
    where $\boldsymbol{x} \in \mathbb{R}^p$ is a $p$-dimensional covariate vector, $t \in \mathbb{R}$ is an additional predictor entering through an unknown smooth function $g(\cdot)$, and $\varepsilon$ is an error term. We rewrite the nonparametric elements in model \eqref{eq:1} as

    \begin{equation}
    	Y = \boldsymbol{x}^\top \boldsymbol{\beta} + \boldsymbol{n}^\top \boldsymbol{g} + \varepsilon,
    	\label{eq:2}
    \end{equation}
    where $\boldsymbol{n}$ is a $q \times 1$ incidence vector with the $j$-th element equal to the indicator function $I(t = t_m^0)$ for $m = 1, \ldots, q$, and $\mathbf{g} = (g(t_1^0), \ldots, g(t_q^0))$ with $t_1^0, \ldots, t_q^0$ being the distinct and ordered values of $t$ (see e.g., \cite{Bazrafkan}). Building on this idea, \cite{Wu,Skhosana,Hwang} extended the finite mixture of regressions framework to incorporate partially linear structures within heterogeneous populations.
    
    In many practical settings, researchers must contend with censored and often missing observations. Such censoring commonly arises due to detection limits of measuring devices or restrictions built into study designs. For example, clinical trials, economic data collections, and environmental monitoring programs frequently exhibit left- or right-censoring when the actual values lie beyond the range that can be reliably recorded. In mixture models, where data come from several hidden subpopulations censored data (partially observed values) and missing data (unobserved values) make it harder to identify each component. Methods like the EM algorithm handle this by treating the unobserved parts as latent variables and adjusting the likelihood accordingly.
    In this work, we introduce a more adaptable and resilient framework: a mixture of censored partially linear expert models built upon the two–piece scale mixture of normal (TP–SMN) distributions (\cite{Maleki}). The TP–SMN family extends traditional SMN models by allowing asymmetry, making it a powerful tool for capturing data features such as skewness and heavy–tailed behavior. Compared with symmetric scale mixture of normal (SMN; \cite{Andrews}) or asymmetric scale mixture of skew-normal (SMSN; \cite{Branco}) formulations, TP–SMN distributions offer enhanced flexibility, which is especially valuable when modelling censored outcomes that may include outliers or asymmetric patterns. One of the main strengths of the proposed finite mixture of censored partially linear expert models built on TP–SMN distributions is that all conditional expectations needed in the E–step of the ECME algorithm can be obtained in closed form. This removes the need for MCMC–based approximations, which are often computationally intensive and prone to numerical inaccuracy, and consequently yields more stable and precise maximum likelihood estimates compared with other mixture models employing skewed distributions. Overall, the proposed approach is well suited for applications where censoring and departures from normality occur simultaneously. Extensive simulation studies and analyses of real datasets illustrate the effectiveness and practicality of the proposed semiparametric mixture censored partially linear expert framework.
    The remainder of this paper is organized as follows: Section \ref{sec.2} reviews relevant properties of the TP-SMN family and the associated components needed for an EM–type algorithm to obtain maximum penalized likelihood (MPL) estimates. Section \ref{sec.3} introduces the semiparametric mixture of censored partially linear expert (MoCPLE) models based on the TP–SMN family and outlines the EM-type algorithm for parameter estimation along with the observed information matrix. Section \ref{sec.4} presents simulation results and real data analyses to evaluate the performance of the proposed method. Finally, Section \ref{sec.5} concludes with some remarks and future research directions.

	\section{The TP-SMN class of distributions (An overview)}\label{sec.2}
	In this section, we review some properties of the TP-SMN family of distributions.The well known SMN family introduced by Andrews and Mallows (1974) has pdf
	\begin{equation}
		f_{\mathrm{smn}}(x; \mu,\sigma,\bm{\nu})
		=
		\int_0^\infty
		\phi(x; \mu, u^{-1}\sigma^2)
		\, dH(u; \bm{\nu}),
		\quad x \in \mathbb{R},
		\label{eq:3}
	\end{equation}
	where $\phi(\cdot; \mu, \sigma^{2})$ represents the density of the normal distribution 
	$\mathcal{N}(\mu,\sigma^{2})$, and $\mathcal{H}(\cdot; \bm{\nu})$ plays the role of an arbitrary cumulative 
	distribution function (cdf) containing $\bm{\nu}$, which includes several members and is denoted by 
	$X \sim \mathcal{SMN}(\mu, \sigma, \bm{\nu})$.
	
	The TP-SMN is a rich class of distributions constructed based on the SMN family that covers 
	symmetric/asymmetric and light/heavy-tailed distributions in terms of their pdf and cdf. 
	For $x \in \mathbb{R}$, this family is represented as
	
	\begin{equation}
		f(x; \mu, \sigma, \gamma, \bm{\nu}) =
		\begin{cases}
			2(1 - \gamma)\,
			f_{\mathrm{smn}}\!\left(x; \mu, \sigma(1 - \gamma), \bm{\nu}\right), & x \le \mu, \\[6pt]
			2\gamma\,
			f_{\mathrm{smn}}\!\left(x; \mu, \sigma\gamma, \bm{\nu}\right), & x > \mu,
		\end{cases}
		\label{eq:4}
	\end{equation}	
	\begin{equation}
		F(x; \mu, \sigma, \gamma, \bm{\nu}) =
		\begin{cases}
			2(1 - \gamma)\,
			F_{\mathrm{smn}}\!\left(x; \mu, \sigma(1 - \gamma), \bm{\nu}\right), & x \le \mu, \\[6pt]
			1 - 2\gamma + 2\gamma\,
			F_{\mathrm{smn}}\!\left(x; \mu, \sigma\gamma, \bm{\nu}\right), & x > \mu,
		\end{cases}
		\label{eq:5}
	\end{equation}

	\medskip
	\noindent
	where $0 < \gamma < 1$ is the slant parameter, 
	$f_{\mathrm{smn}}(\cdot;\cdot)$ and 
	$F_{\mathrm{smn}}(\cdot;\cdot)$ denote the pdf and cdf of the SMN distribution, respectively.
	A random variable with pdf and cdf given in \eqref{eq:4} and \eqref{eq:5} is denoted by 
	$X \sim \mathcal{TP\text{-}SMN}(\mu, \sigma, \gamma, \bm{\nu})$.
	which has the following stochastic representation:
	\begin{equation}
		X = \mu + F U^{-1/2} |Z|,\qquad Z\perp U \perp F,
		\label{eq:6}
	\end{equation}
	where $Z \sim \mathcal {N}(0, \sigma^2)$, the scalar random variable $U \sim \mathcal {H}(\cdot; \bm{\nu})$ and the shape random variable $F$, with the following probability mass function (pmf):
	\[
	P(F=f) = \gamma^{\frac{1+s}{2}} (1 - \gamma)^{\frac{1-s}{2}}, \quad \text{where } \quad s = \mathrm{Sign}(f), \quad f \in \{\gamma, -(1-\gamma)\}.
	\]
	
	\noindent
	The advantage of the above representation of the binary random variable $F$ becomes evident in the construction of the penalized log-likelihood function of the model.
	
	To illustrate the richness of the $\mathrm{TP\text{-}SMN}$ class of distributions, note that various members of this family can be obtained by choosing different distributions for the scale-mixing random variable $U$ in \eqref{eq:6}. Specifically, we have:
	\begin{itemize}
		\item \textbf{Two-piece Normal (TP–N):} $U=1$ with probability one.
		\item \textbf{Two-piece t (TP–T):} $U \sim \mathrm{Gamma}(\nu/2, \nu/2)$, $\nu>0$.
		\item \textbf{Two-piece Slash (TP–SL):} $U \sim \mathrm{Beta}(\nu, 1)$, $\nu>0$.
		\item \textbf{Two-piece Contaminated-Normal (TP–CN):} $P(U=u) = \nu \mathbb{I}_{(u=\tau)} + (1-\nu) \mathbb{I}_{(u=1)}$; $0 < \nu < 1$, $0 < \delta \le 1$.
	\end{itemize}
	
	\noindent
	Finally, the proposed $\mathrm{TP\text{-}SMN}$ random variable $X$ has mean and variance given by:
	\[
	E(X) = \mu - b\Delta \quad \text{and} \quad \mathrm{Var}(X) = \sigma^2 [c_2 k_2 - b^2 c_1^2],
	\]
	where $b = \sqrt{2/\pi} k_1$, $\Delta = \sigma(1 - 2\gamma)$, $c_r = \gamma^{r+1} + (-1)^r (1 - \gamma)^{r+1}$ and $k_r = E(U^{-r/2})$, $r \in \mathbb{Z}$.
	
	\begin{remark}\label{rem1} 
		Let $X \sim \mathcal{TP\text{-}SMN}(\mu, \sigma, \gamma, \bm{\nu})$ and $A=(a,b)$ for which $a < b$ such that $P(X \in A) > 0$. A random variable $X$ has a truncated two-piece scale mixtures of normal distributions on the interval $A$ if it has the same distribution as $Y = X \mid X \in A$ and is denoted by $Y \sim \mathcal{TTP\text{-}SMN}(\mu, \sigma, \gamma, \bm{\nu}; A)$ with the following pdf:
		\[
		f_{T}(y; \mu, \sigma, \gamma, \bm{\nu}; A) =
		\begin{cases}
			\dfrac{f(y; \mu, \sigma, \gamma, \bm{\nu})}{F(b; \mu, \sigma, \gamma, \bm{\nu}) - F(a; \mu, \sigma, \gamma, \bm{\nu})}, & y \in A, \\[10pt]
			0, & y \notin A,
		\end{cases}
		\]
		where $f(\cdot; \cdot)$ and $F(\cdot; \cdot)$ are the pdf and cdf of the $\mathrm{TP\text{-}SMN}$ distributions, respectively, given in \eqref{eq:4} and \eqref{eq:5}.
	\end{remark}
	The essential conditional expectations for the non-truncated and interval-truncated $\mathrm{TP\text{-}SMN}$ family used for parameter estimation are provided in the Appendix, with further details available in \cite{Maleki}.
	
	\section{The TP–SMN Mixture of Censored Partially Linear Expert Model}\label{sec.3}
	Our focus is on semiparametric mixtures of partially linear models, where the responses may be observed or censored and the errors follow independent and identically distributed (i.i.d.) $\mathrm{TP\text{-}SMN}$ distributions. So we introduce a semiparametric mixture of censored partially linear experts model which hereafter so called $\mathrm{TP\text{-}SMN\text{-}MoCPLE}$ model.
	
	\subsection{Model specification}\label{sub.3.1}
	Mixtures of partially linear experts ($\mathrm{MoPLE}$) models are constructed by treating each expert network of the $\mathrm{MoE}$ model, for $k=1,\ldots,K$ sub-populations, as a partial linear model \eqref{eq:2}. The expert components are defined using $\mathrm{TP\text{-}SMN}$ distributions and specifying the component means via polynomial regression and the mixing proportions via a multinomial logistic function, the resulting pdf can be expressed as:
	\begin{equation}
		\label{eq:7}
		f_{Y}(y_i; \bm{x}_i, \mathbf{\Theta}) = \sum_{k=1}^{K} \pi_k(\bm{r}_i; \bm{\tau}) \, f_k\left(y_i; \bm{x}_i^{\top} \boldsymbol{\beta}_k + \bm{n}_i^{\top} \mathbf{g}_k, \sigma_k, \gamma_k, \bm{\nu}_k \right); \quad i=1,\ldots,n,
	\end{equation}	
	where $\mathbf{Y}=(Y_1,\ldots,Y_n )^{\top}$ is the vector of response variables, $\bm{x}_i \in \mathbb{R}^p$, and $\bm{r}_i=(1,r_1,\ldots,r_{d-1} )^{\top} \in \mathbb{R}^d$ are the vector of explanatory, additional predictor and covariate variables corresponding to $Y_i$ for $i=1,\ldots,n$, $f_k(\cdot \mid \cdot)$ is the $\mathrm{TP\text{-}SMN}$ pdf \eqref{eq:4}, and for the gating parameters $\bm{\tau}=(\mathbf{\tau}_1^{\top},\ldots, \mathbf{\tau}_{K-1}^{\top} )^{\top}$ with $\bm{\tau}_k=(\tau_{k0},\ldots, \tau_{k(d-1)} )^{\top}$, and
	\[
	\pi_k (\bm{r}; \bm{\tau}) = \frac{\exp(\bm{\tau}_k^{\top} \bm{r})}{1 + \sum_{k'=1}^{K-1}\exp(\bm{\tau}_{k'}^{\top} \bm{r})}, \quad k=1,\ldots,K,
	\]
	for which the model’s parameter set is $\mathbf{\Theta}=(\mathbf{\Theta}_1^{\top},\ldots, \mathbf{\Theta}_K^{\top}, \bm{\tau})^{\top}$ such that $\mathbf{\Theta}_k=(\boldsymbol{\beta}_k^{\top}, \mathbf{g}_k^{\top}, \sigma_k, \gamma_k, \bm{\nu}_k ), \quad k=1,\ldots,K$.
	
	\medskip	
	In Mixture of Experts (MoE) models, the gating network utilizes a Softmax function to assign weights to experts. Since the Softmax output is invariant to the addition of a constant vector $\bm{c}$ to all expert parameters (i.e., $\bm{\tau}_k \rightarrow \bm{\tau}_k + \bm{c}$ yields identical probabilities), the model parameters become non-identifiable, leading to infinite equivalent solutions. To resolve this redundancy and ensure a unique parameter estimation during optimization, we constrain the parameter vector of the first expert to zero, setting $\bm{\tau}_1 = 0$ as a reference baseline. Furthermore, along with this identifiability constraint on the gating parameters, assuming that the smooth functions within each expert are continuous, the covariate $\bm{x}_i$ is non-constant and not a deterministic function of $t$, and the parameter sets across experts are distinct, ensures the overall identifiability of the proposed model.
	
	\medskip
	In the proposed $\mathrm{TP\text{-}SMN\text{-}MoPLE}$ models, we assume also that the response variables are censored observations and called it by $\mathrm{TP\text{-}SMN}$ mixture of censored partially linear experts ($\mathrm{MoCPLE}$) models. In other words, some of the responses are subject to censoring, which may take the form of interval-, left-, or right-censoring. Thus, let the available response variable $Y_i$ be presented as the joint variables $(\mathbf{W}_i, \rho_i)$, where $\mathbf{W}_i$ represents the uncensored reading ($\mathbf{W}_i=Y_{io}$ ) or interval censoring ($\mathbf{W}_i=(C_{i1},C_{i2})$ ), and $\rho_i$ is the censoring indicator as:
	\[
	\rho_i = \begin{cases}
		1, & \text{if } C_{i1} \le Y_i \le C_{i2} \\
		0, & \text{if } Y_i = Y_{i0}
	\end{cases}
	\]
	Note that in this setting, if $C_{i1}=-\infty$ (or $C_{i2}=+\infty$), the left-censoring (or right-censoring) is occurred, if $C_{i1}=-\infty$ and $C_{i2}=+\infty$, the missing is occurred, and in the case $-\infty < C_{i1} < C_{i2} < +\infty$ the interval-censored realization is observed. We establish our methodology based on the interval-censoring scheme; however, the left/right-censoring and missing schemes are also investigated in the simulation and real-data analyses.
	Under this framework, $\mathbf{Y}$ is partitioned into observed values and censored observations. Therefore, $\mathbf{Y}$ is treated as a latent variable due to its partial observability. Given these assumptions and using Remark \ref{rem1}, the penalized log-likelihood function of the $\mathrm{MoCPLE}$ model can be written as
	\begin{eqnarray} \label{eq:8}
		\begin{split}
			\ell_p (\bm{\Theta},\bm{\lambda}| \mathbf{w},\bm{\rho}) &= \sum_{i=1}^n \log \sum_{k=1}^K \left\{ \pi_k (\bm{r}_k; \bm{\tau}) \left[ f(w_i; \mu_{ik}, \sigma_k, \gamma_k, \bm{\nu}_k) \right]^{1-\rho_i} \right. \\
			& \quad \left. \left[ F(c_{i2}; \mu_{ik}, \sigma_k, \gamma_k, \bm{\nu}_k) - F(c_{i1}; \mu_{ik}, \sigma_k, \gamma_k, \bm{\nu}_k) \right]^{\rho_i} \right\}\\ 
			& \quad - \sum_{k=1}^K \frac{\lambda_k}{2} J(\textbf{g}_k)
		\end{split},
	\end{eqnarray}
	where $\mu_{ik}=\bm{x}_i^\top \bm{\beta}_k + \bm{n}_i^\top \mathbf{g}_k$, $\bm{\lambda} = (\lambda_1,\dots,\lambda_K)^\top$, $\mathbf{w} = (w_1,\dots,w_n)^\top$ and $\boldsymbol{\rho} = (\rho_1,\dots,\rho_n)^\top$
	denote the realizations of $\mathbf{W} = (W_1,\dots,W_n)^\top$
	and $\boldsymbol{\rho} = (\rho_1,\dots,\rho_n)^\top$, respectively. The term $J(\textbf{g}_k)$ is the penalty function over $\textbf{g}_k(\cdot)$ and $\alpha_k$ is a smoothing parameter that controls the trade-off between goodness-of-fit, measured by large values of $\ell_p (\bm{\Theta},\bm{\lambda}| \mathbf{w},\bm{\rho})$, and smoothness, measured by small values of $J(\textbf{g}_k)$.	Following \cite{Ibacache}, we use a natural cubic spline to estimate $\textbf{g}_k(\cdot)$. The general form of the penalty function is
	\[
	J(\textbf{g}_k)=\int_a^b
	\left[g_k''(t)\right]^2 dt,
	\]
	where $g_k''(t)$ denotes the second  derivative of $g_k(t)$ on $[a,b]$ which contains the values $t_1^0,\dots,t_q^0$.
	According to Theorem 2.1 of \cite{Green}, $
	\int_a^b
	\left[g_k''(t)\right]^2 dt
	=
	g_k^\top \bm{L} g_k,
	$
	for which $\bm{L} \in \mathbb{R}^{q\times q}$ is a non–negative definite matrix depending only on the knot differences.
	
	\medskip
	The maximum penalized likelihood (MPL) estimates are obtained by maximizing the penalized log-likelihood. Because the log-likelihood function in \eqref{eq:8} is analytically intractable, closed-form MPL estimates are unavailable, and numerical optimization is required. By exploiting the stochastic representation in \eqref{eq:6} together with the hierarchical representation of the TP-SMN distribution, we develop an expectation conditional maximization either (ECME; \cite{Liu}) algorithm to estimate parameters from the TP-SMN-MoCPLE models. The ECME algorithm is an extension of the expectation conditional maximization (ECM; \cite{Meng}) that not only inherits its stable properties (e.g., monotone convergence and implementation simplicity) but also can be faster than ECM. The iterative ECME algorithm replaces some CM-steps of the ECM with CML-steps that maximize the corresponding constrained log-likelihood function instead.
	
	\subsection{MPL estimation of the MoCPLE models parameters via an EM-Algorithm}\label{sub.3.2}
	Beginning with pdf \eqref{eq:7}, we define the component label vector 
	\[
	\bm{Z}_i = (Z_{i1},\dots,Z_{iK})^\top,
	\quad i=1,\dots,n,
	\]
	such that $Z_{ik}=1$ if and only if $Z_i^* = k$. Under this formulation, we have
	\[
	Y_i \mid (Z_{ik}=1)
	\sim
	\mathcal{TP-SMN}
	\left(
	\bm{x}_i^\top \bm{\beta}_k + \bm{n}_i^\top \mathbf{g}_k,
	\sigma_k,
	\gamma_k,
	\bm{\nu}_k
	\right).
	\]
	and under the above approach, the latent random vector $Z_i,i=1,…,n$, has the following multinomial pmf:
	
	\begin{equation}
		P(Z_i=z_i)
		=
		\prod_{k=1}^{K}
		\pi_k(\bm{r}_i;\bm{\tau})^{z_{ik}},
		\quad
		\sum_{k=1}^{K}\pi_k(\bm{r}_i;\bm{\tau})=1.
		\label{eq:9}
	\end{equation}
	\begin{remark}\label{rem2}
		Using the proposed stochastic representation, we obtain
		\begin{align}
			&Y_i \mid U_i=u_i, F_i=f_i, Z_{ik}=1\sim
			\mathcal{TN}
			\left(
			\mu_{ik},
			u_i^{-1} f_i^2 \sigma_k^2
			\right)
			\mathbb{I}_B(y_i),
			\nonumber\\
			&U_i \mid Z_{ik}=1\sim \mathcal{H}(u_i;\bm{\nu}_k),
			\nonumber\\
			&Z_i \mid \bm{r}_i \sim
			\mathcal{M}
			(1;\pi_1(\bm{r}_i;\bm{\tau}),\dots,\pi_K(\bm{r}_i;\bm{\tau})),
			\nonumber\\
			&P(F_i=f_i \mid Z_{ik}=1)=
			\gamma_k^{\frac{1+s_i}{2}}
			(1-\gamma_k)^{\frac{1-s_i}{2}},
			\quad f_i=\gamma_k,-(1-\gamma_k),
			\label{eq:10}
		\end{align}
		
		\noindent
		for $i=1,…,n, k=1,…,K$ , where $s_i=\mathrm{sign}(f_i)$, $\mathcal{TN}(\cdot)\mathbb{I}_B(\cdot)$ denotes the truncated normal distribution on the interval $B$, for which $B=(-\infty, \bm{x}_i^\top \bm{\beta}_k + \bm{n}_i^\top \mathbf{g}_k]$ if $f_i < 0$, and $B=(\bm{x}_i^\top \bm{\beta}_k + \bm{n}_i^\top \mathbf{g}_k, +\infty)$ if $f_i \geq 0$ and $\mathcal{M}(1;\cdot)$ denotes the one trail multinomial distribution.
	\end{remark}
	
	\bigskip
	Let
	$\mathbf{y}_{\text{com}}
	=
	(\mathbf{w}^\top,
	\boldsymbol{\rho}^\top,
	\mathbf{y}^\top,
	\mathbf{r}^\top,
	\mathbf{u}^\top,
	\mathbf{f}^\top,
	\mathbf{z}^\top)^\top
	$, then $
	\ell_{cp}(\bm{\Theta})
	=
	\sum_{i=1}^{n}
	\ell_{icp}(\bm{\Theta})
	$,
	where
	\begin{align}
		\ell_{icp}(\bm{\Theta})
		=
		\sum_{k=1}^{K}
		z_{ik}
		\Bigg\{
		&
		\log \pi_k(\bm{r}_i;\bm{\tau})
		-
		\log \sigma_k
		-
		\frac{u_i}{2 f_i^2 \sigma_k^2}
		(y_i-\mu_{ik})^2
		\nonumber\\
		&
		+
		\frac{1+s_i}{2}\log\gamma_k
		+
		\frac{1-s_i}{2}\log(1-\gamma_k)
		+
		\log h(u_i;\bm{\nu}_k)
		-
		\frac{\lambda_k}{2}J(\textbf{g}_k)
		\Bigg\}.
		\label{eq:11}
	\end{align}
	
	Let the superscript ($m$) indicates the estimate of the related parameter at $m$ stage of the algorithm. In the E-step of the algorithm, we must obtain the so-called Q-function 
	$
	Q(\bm{\Theta}| \bm{\Theta}^{(m)})
	=
	\mathbb{E}
	\left[
	\ell_{cp}(\bm{\Theta})
	\mid
	\mathbf{w},\boldsymbol{\rho},\bm{\Theta}^{(m)}
	\right]
	$
	is given by:
	\begin{align}
		Q(\bm{\Theta}| \bm{\Theta}^{(m)})
		&=
		\sum_{i=1}^{n}
		\sum_{k=1}^{K}
		\hat z_{ik}^{(m)}
		\Bigg\{
		\log \pi_k(\bm{r}_i;\bm{\tau})
		-\frac{1}{2}\log \sigma_k
		\nonumber\\
		&\quad
		-\frac{1}{2\sigma_k^2}
		\Big(
		\hat\xi_{2,ik}^{(m)}
		-2(\bm{x}_i^\top \bm{\beta}_k + \bm{n}_i^\top \mathbf{g}_k)\hat\xi_{1,ik}^{(m)}
		+(\bm{x}_i^\top \bm{\beta}_k + \bm{n}_i^\top \mathbf{g}_k)^2\hat\xi_{0,ik}^{(m)}
		\Big)
		\nonumber\\
		&\quad
		+\frac{1+\hat s_{ik}^{(m)}}{2}\log\gamma_k
		+\frac{1-\hat s_{ik}^{(m)}}{2}\log(1-\gamma_k)
		+\hat\Upsilon_{ik}^{(m)}
		-\frac{\lambda_k}{2}J(\textbf{g}_k)
		\Bigg\}
		\label{eq:12},
	\end{align}
	where
	\begin{align*}
		\hat z_{ik}^{(m)}
		&=
		\mathbb{E}[Z_{ik}\mid w_i,\rho_i,\hat{\bm{\Theta}}_k^{(m)}],
		\\
		\hat\xi_{j,ik}^{(m)}
		&=
		\mathbb{E}[U_i F_i^{-2} Y_i^j
		\mid w_i,\rho_i,\hat{\bm{\Theta}}_k^{(m)}],
		\quad j=0,1,2,
		\\
		\hat s_{ik}^{(m)}
		&=
		\mathbb{E}[s_i\mid w_i,\rho_i,\hat{\bm{\Theta}}_k^{(m)}],
		\\
		\hat\Upsilon_{ik}^{(m)}
		&=
		\mathbb{E}[\log h(u_i;\bm{\nu}_k)
		\mid w_i,\rho_i,\hat{\bm{\Theta}}_k^{(m)}].
	\end{align*}
	In what follows, we derive expressions for conditional expectations under both uncensored and censored scenarios,
	
	\medskip
	\textbf{Uncensored case ($\rho_i=0$):}
	\begin{align*}
		\hat z_{ik}^{(m)}
		&=
		\frac{
			\pi_k(\bm{r}_i;\hat{\bm{\tau}}^{(m)})
			f_k(y_i; \hat\mu_{ik}^{(m)},\hat\sigma_k^{(m)},\hat\gamma_k^{(m)},\hat{\bm{\nu}}_k^{(m)})
		}{
			\sum_{k'=1}^{K}
			\pi_{k'}(\bm{r}_i;\hat{\bm{\tau}}^{(m)})
			f_{k'}(y_i; \hat\mu_{ik'}^{(m)},\hat\sigma_{k'}^{(m)},\hat\gamma_{k'}^{(m)},\hat{\bm{\nu}}_{k'}^{(m)})
		},
		\\
		\hat\xi_{j,ik}^{(m)}
		&=
		y_i^j
		\mathbb{E}[U_i F_i^{-2}
		\mid Y_i=y_i,\hat{\bm{\Theta}}_k^{(m)}],
		\quad j=0,1,2,
		\\
		\hat s_{ik}^{(m)}
		&=
		\mathbb{E}[S_i\mid Y_i=y_i,\hat{\bm{\Theta}}_k^{(m)}],
		\\
		\hat\Upsilon_{ik}^{(m)}
		&=
		\mathbb{E}[\log h(u_i;\bm{\nu}_k)
		\mid Y_i=y_i,\hat{\bm{\Theta}}_k^{(m)}].
	\end{align*}
	
	\medskip
	\textbf{Censored case ($\rho_i=1$):}
	\begin{align*}
		\hat z_{ik}^{(m)}
		&=
		\frac{
			\pi_k(\bm{r}_i;\hat{\bm{\tau}}^{(m)})
			\big[
			F_k(c_{i2}; \hat\mu_{ik}^{(m)},\hat\sigma_k^{(m)},\hat\gamma_k^{(m)},\hat{\bm{\nu}}_k^{(m)})
			-
			F_k(c_{i1}; \hat\mu_{ik}^{(m)},\hat\sigma_k^{(m)},\hat\gamma_k^{(m)},\hat{\bm{\nu}}_k^{(m)})
			\big]
		}{
			\sum_{k'=1}^{K}
			\pi_{k'}(\bm{r}_i;\hat{\bm{\tau}}^{(m)})
			\big[
			F_{k'}(c_{i2}; \hat\mu_{ik'}^{(m)},\hat\sigma_{k'}^{(m)},\hat\gamma_{k'}^{(m)},\hat{\bm{\nu}}_{k'}^{(m)})
			-
			F_{k'}(c_{i1}; \hat\mu_{ik'}^{(m)},\hat\sigma_{k'}^{(m)},\hat\gamma_{k'}^{(m)},\hat{\bm{\nu}}_{k'}^{(m)})
			\big]
		},
		\\
		\hat\xi_{j,ik}^{(m)}
		&=
		\mathbb{E}[U_i F_i^{-2} Y_i^j
		\mid c_{i1}\le Y_i\le c_{i2},\hat{\bm{\Theta}}_k^{(m)}],
		\quad j=0,1,2,
		\\
		\hat s_{ik}^{(m)}
		&=
		\mathbb{E}[S_i
		\mid c_{i1}\le Y_i\le c_{i2},\hat{\bm{\Theta}}_k^{(m)}],
		\\
		\hat\Upsilon_{ik}^{(m)}
		&=
		\mathbb{E}[\log h(u_i;\bm{\nu}_k)
		\mid c_{i1}\le Y_i\le c_{i2},\hat{\bm{\Theta}}_k^{(m)}].
	\end{align*}
	
	where $\hat\mu_{ik}^{(m)}
	=
	\bm{x}_i^\top \hat{\bm{\beta}}_k^{(m)}
	+
	\bm{n}_i^\top \hat {\bm{g}}_k^{(m)}$.
	
	\medskip
	\noindent
	Computing $\hat\xi_{j,ik}^{(m)}$, $j=0,1,2$, and $\hat s_{ik}^{(m)}$ for uncensored and censored responses are obtained via Lemmas \ref{lem:A1},\ref{lem:A2} and \ref{lem:A3} in the \ref{appa}. 
	
	\begin{itemize}
		\item {\bf E-step:}
		Knowing $\bm{\Theta}=\hat{\bm{\Theta}}^{(m)}$ and using the above calculations, we compute the corresponding conditional expectations. Then we maximize $Q(\bm{\Theta}| \hat{\bm{\Theta}}^{(m)})$ in the EM approach by solving the first–order condition
		\[
		\nabla Q(\bm{\Theta}| \hat{\bm{\Theta}}^{(m)})=\bm{0},
		\]
		where $\nabla$ denotes the gradient operator, yielding the following updates.
		
		\medskip
		\item {\bf CM-step:}
		Update $\bm{\Theta}^{(m)}$ (denoted by $\bm{\Theta}^{(m+1)}$) by maximizing 
		$Q(\bm{\Theta}| \bm{\Theta}^{(m)})$ over $\bm{\Theta}$. For $k=1,\dots,K$ we obtain:
		\item {\bf CM-step 1:} According to the proposition 2 of \cite{Nguyen}, the update of $\bm\tau_k$ can be update as
		
		\begin{align*}
			\hat{\bm{\tau}}_k^{(m+1)}
			&=
			4\left(
			\sum_{i=1}^{n} \bm{r}_i \bm{r}_i^\top
			\right)^{-1}
			\left(
			\sum_{i=1}^{n}
			\big(
			\hat z_{ik}^{(m)}
			-\pi_k(\bm{r}_i;\hat{\bm{\tau}}_k^{(m)})
			\big)
			\bm{r}_i
			\right)
			+\hat{\bm{\tau}}_k^{(m)},
		\end{align*} 
		\item {\bf CM-step 2:}
		\begin{align*} 
			\hat{\bm{\beta}}_k^{(m+1)}
			&=
			\left(
			\sum_{i=1}^{n}
			\hat z_{ik}^{(m)}
			\hat\xi_{0,ik}^{(m)}
			\bm{x}_i \bm{x}_i^\top
			\right)^{-1}
			\sum_{i=1}^{n}
			\hat z_{ik}^{(m)}
			\Big(
			\hat\xi_{1,ik}^{(m)}
			-
			\hat\xi_{0,ik}^{(m)}
			\bm{n}_i^\top \hat{\bm{g}}_k^{(m)}
			\Big)
			\bm{x}_i,
			\\[6pt]
			\hat{\bm{g}}_k^{(m+1)}
			&=
			\left(
			\sum_{i=1}^{n}
			\hat z_{ik}^{(m)}
			\Big(
			\hat\xi_{0,ik}^{(m)} \bm{n}_i \bm{n}_i^\top
			+
			\lambda_k \hat\sigma_k^{2(m)} \bm{L}
			\Big)
			\right)^{-1}
			\sum_{i=1}^{n}
			\hat z_{ik}^{(m)}
			\Big(
			\hat\xi_{1,ik}^{(m)}
			-
			\hat\xi_{0,ik}^{(m)}
			\bm{x}_i^\top \hat{\bm{\beta}}_k^{(m+1)}
			\Big)
			\bm{n}_i,
			\nonumber
			\\[6pt]
			\hat\sigma_k^{2(m+1)}
			&=
			\left(
			\sum_{i=1}^{n}
			\hat z_{ik}^{(m)}
			\right)^{-1}
			\sum_{i=1}^{n}
			\hat z_{ik}^{(m)}
			\Big[
			\hat\xi_{0,ik}^{(m)}
			\mu_{ik}^{2(m+1)}-
			2\hat\xi_{1,ik}^{(m)}
			\mu_{ik}^{(m+1)}
			+
			\hat\xi_{2,ik}^{(m)}
			\Big],
			\nonumber
			\\[6pt]
			\hat\gamma_k^{(m+1)}
			&=
			\frac{1}{2}
			\left(
			1+
			\frac{
				\sum_{i=1}^{n}
				\hat z_{ik}^{(m)}
				\hat \zeta_{ik}^{(m)}
			}{
				\sum_{i=1}^{n}
				\hat z_{ik}^{(m)}
			}
			\right),
		\end{align*}
		where
		$\hat\mu_{ik}^{(m+1)}
		=
		\bm{x}_i^\top \hat{\bm{\beta}}_k^{(m+1)}
		+
		\bm{n}_i^\top \hat{\bm{g}}_k^{(m+1)}$.
		
		\medskip
		
		\item {\bf CML-step:} Update $\bm{\nu} = (\nu_1^\top,\dots,\nu_K^\top)^\top$by maximizing the actual penalized log-likelihood:
		
		\begin{equation}
			\hat{\bm{\nu}}^{(m+1)}
			=
			\arg\max_{\nu}
			\ell_p\big(
			\bm{\Theta}_{(-\bm{\nu})}^{(m+1)}, \bm{\lambda}
			\mid \mathbf{w},\boldsymbol{\rho}
			\big).
			\label{eq:13}
		\end{equation}
		where $\ell_p(\cdot| \mathbf{w},\boldsymbol{\rho})$ is defined in \eqref{eq:8} and 
		$\bm{\Theta}_{(-\bm{\nu})}^{(m+1)}$ denotes the updated parameter vector without $\bm{\nu}$ at iteration $m+1$.
	\end{itemize}

	To start the algorithm, initial values are required, so First, the response data are partitioned into $K$ groups using the $k$-means clustering algorithm. The proportion of observations belonging to each group is considered as the initial mixing weights (component probabilities). Then, the initial values of the regression coefficients for each component are obtained by applying the least squares method within each corresponding partitioned group. Finally, based on the computed residuals, the initial values of the parameters of the TP-SMN family are obtained using the procedure proposed by \cite{Maleki}. 
	The algorithm iterates between the E-step and the CML-step until convergence is achieved. Convergence is assessed by monitoring the difference between two successive evaluations of the log-likelihood function. Specifically, the algorithm stops when
	$
	\left|
	\ell^{(m+1)} - \ell_\infty^{(m+1)}
	\right|
	< \varepsilon,
	$
	where $\varepsilon$ is a predetermined tolerance level. Here, $\ell^{(m+1)}$ denotes the value of the log-likelihood evaluated at $\bm{\Theta}^{(m+1)}$, and $\ell_\infty^{(m+1)}$ is the asymptotic estimate of the log-likelihood at the $(m+1)$th iteration, computed using Aitken’s acceleration as
	\[
	\ell_\infty^{(m+1)}
	=
	\ell^{(m)}
	+
	\frac{
		\ell^{(m+1)} - \ell^{(m)}
	}{
		1 - c^{(m)}
	}.
	\]
	where the Aitken acceleration at iteration $m$ is defined as $c^{(m)}
	=
	\frac{
		\ell^{(m+1)} - \ell^{(m)}
	}{
		\ell^{(m)} - \ell^{(m-1)}
	}$.
	
	\medskip
	To address the label switching problem and ensure parameter identifiability, we impose an ordering constraint on the gating parameters. Since the parameter vectors $\bm \tau_k$ are multidimensional, we define a fixed reference vector $\bm{a}$ to project them into scalars. A common choice is the vector of ones, $\bm{a} = (1, \dots, 1)^\top$, in which case $\bm\tau_k^\top \bm{a}$ represents the sum of the elements of $\bm\tau_k$. We then constrain the parameters such that $\bm\tau_1=\bm{0}$ and $\bm\tau_2^\top \bm{a} < \dots < \bm\tau_K^\top \bm{a}$. This constraint eliminates the symmetry of the likelihood function and guarantees a unique solution during the estimation process.
	
	\subsection{Effective degrees of freedom and model selection}\label{sub.3.3}
	
	In additive models, the Akaike information criterion (AIC; \cite{Akaike}) can be applied to select an appropriate $\alpha_k$, $k=1,\dots,K$. The AIC for MoCPLE models are given by
	
	\[
	\mathrm{df}(\lambda_k)
	=
	\operatorname{tr}
	\left\{
	\bm{N} (\bm{N}^\top \bm{N} + \lambda_k \hat\sigma_k^2 \bm{L})^{-1} \bm{N}^\top
	\right\}
	=
	\operatorname{tr}
	\left\{
	\bm{N}^\top \bm{N} (\bm{N}^\top \bm{N} + \lambda_k \hat\sigma_k^2 \bm{L})^{-1}
	\right\}.
	\]
	Let $\bm{Q} = \bm{N}^\top \bm{N}$ and $
	\bm{E}_k = \hat\sigma_k^2 \bm{Q}^{-1/2} \bm{L} \bm{Q}^{-1/2}$
	, then, from \cite{Eilers}, we may express
	
	\[
	\mathrm{df}(\lambda_k)
	=
	\operatorname{tr}
	\left\{
	\bm{Q}^{1/2}
	\left(
	\bm{Q}^{1/2} \bm{Q}^{1/2} + \lambda_k \hat\sigma_k^2 \bm{L}
	\right)^{-1}
	\bm{Q}^{1/2}
	\right\}
	=
	\operatorname{tr}
	\left\{
	(\bm{I}_q + \lambda_k \bm{E}_k)^{-1}
	\right\}
	=
	\sum_{j=1}^{q}
	\frac{1}{1 + \lambda_k e_{kj}},
	\]
	where $e_{kj}$ is the $j$th eigenvalue of $\bm{E}_k$ and $\bm{I}_q$ is the identity matrix of order $q$. Thus, it follows that $\mathrm{df}(\lambda_k)$ is inversely proportional to $\lambda_k$. Therefore, one has a total of $p + \psi + \mathrm{df}(\lambda_k)$ parameters to be estimated, for which $p$ is the dimension of the regression parameters, and $\psi$ is the number of parameters of the proposed TP-SMN distributions in the $k$-th component. The Akaike information criterion is useful for selecting an appropriate model for the given data with adequate sample size, and applied in minimizing the function given by
	
	\[
	\mathrm{AIC}(\lambda_k)
	=
	-2 \ell_p(\hat{\bm{\Theta}}, \lambda_k| \bm{w}, \bm{\rho})
	+
	2 \{ p + \psi + \mathrm{df}(\lambda_k) \},
	\]
	where $\ell_p(\hat{\bm{\Theta}}, \lambda_k| \bm{w}, \bm{\rho})$ is the component-wise of the penalized log-likelihood function \eqref{eq:8} available at $\hat{\bm{\Theta}}$ for a fixed $\lambda_k$. We employ a grid search strategy over the $\lambda$-space to determine the optimal value of the penalty parameter $\bm{\lambda}=(\lambda_1,\dots,\lambda_K)^\top$.
	
	\subsection{Standard error approximation}\label{sub.3.4}
	In this section, we obtain the standard errors of the MPL estimates of the parameters of the proposed TP-SMN-CPLE models. Under the usual regularity conditions, we have that the asymptotic covariance matrix of $\bm\Theta$ can be approximated by the inverse of the empirical information matrix defined as
	
	\[
	\bm{I}_0
	=
	\sum_{i=1}^{n}
	\bm{s}(y_i \mid \bm{\Theta})
	\bm{s}^\top(y_i| \bm{\Theta})
	-
	n^{-1}
	\left(
	\sum_{i=1}^{n}
	\bm{s}(y_i| \bm{\Theta})
	\right)
	\left(
	\sum_{i=1}^{n}
	\bm{s}(y_i| \bm{\Theta})
	\right)^\top,
	\]
	where
	$
	\bm{s}(y_i| \bm{\Theta})
	=
	\frac{\partial Q_i(\bm{\Theta}| \bm{\Theta}^{(r)})}{\partial \bm{\Theta}},
	$
	and $Q_i(\bm{\Theta}| \bm{\Theta}^{(r)})$ is the summand in \eqref{eq:12}. Substituting $\bm{\Theta}$ by the MPL estimates $\hat{\bm{\Theta}}$ in $\bm{I}_0$, conclude its approximation in the form of
	$
	\hat {\bm{I}}_0
	=
	\sum_{i=1}^{n}
	\hat {\bm{s}_i} \hat {\bm{s}}_i^\top,
	$
	for which $\hat {\bm{s}}_i = s(y_i| \hat{\bm{\Theta}})$ is an individual score vector given by
	\[
	\hat {\bm{s}}_i = \biggl( \hat s_{(i,\bm{\tau}_1)},\dots,\hat s_{(i,\bm{\tau}_K)}, \hat s_{(i,\bm{\beta}_1)}^\top,\dots,\hat s_{(i,\bm{\beta}_K)}^\top, \\
	\hat s_{(i,\textbf{g}_1)}^\top,\dots,\hat s_{(i,\textbf{g}_K)}^\top, \hat s_{(i,\sigma_1)},\dots,\hat s_{(i,\sigma_K)},\hat s_{(i,\gamma_1)},\dots,\hat s_{(i,\gamma_K)}, \\
	\hat s_{(i,\bm{\nu}_1)}^\top,\dots,\hat s_{(i,\bm{\nu}_K)}^\top \biggr)
	\]
	with explicit expressions for the elements of $\hat {\bm{s}}_i$ given by
	
	\begin{align}
		&\hat s_{(i,\bm{\tau}_k)}
		=
		\left[
		\hat z_{ik}-\pi_k(\bm{r}_i;\hat{\bm{\tau}})
		\right]\bm{r}_i,
		\nonumber\\
		&\hat s_{(i,\bm{\beta}_k)}
		=
		-\frac{1}{\sigma_k^2}
		\left[
		\hat\xi_{0ik} (\bm{x}_i^\top \hat{\bm{\beta}}_k+\bm{n}_i^\top \hat{\bm{g}}_k)
		-
		\hat{\xi}_{1ik}
		\right]
		\bm{x}_i^\top,
		\nonumber\\
		&\hat s_{(i,\bm{g}_k)}
		=
		-\frac{1}{\sigma_k^2}
		\left[
		\hat\xi_{0ik} (\bm{x}_i^\top \hat{\bm{\beta}}_k+\bm{n}_i^\top \hat{\bm{g}}_k)
		-
		\hat{\xi}_{1ik}
		\right]
		\bm{n}_i^\top-\lambda_k \hat{\bm{g}}_k^\top \bm{L},
		\nonumber\\
		&\hat s_{(i,\sigma_k)}
		=
		-\frac{1}{\hat\sigma_k}
		+
		\frac{1}{\hat\sigma_k^3}
		\left[
		\hat\xi_{0ik} (\bm{x}_i^\top \hat{\bm{\beta}}_k+\bm{n}_i^\top \hat{\bm{g}}_k)^2
		-
		2\hat\xi_{1ik} (\bm{x}_i^\top \hat{\bm{\beta}}_k+\bm{n}_i^\top \hat{\bm{g}}_k)
		+
		\hat\xi_{2ik}
		\right],
		\nonumber\\
		&\hat s_{(i,\gamma_k)}
		=
		\frac{1 + \hat \zeta_{ik}}{2 \hat\gamma_k}
		-
		\frac{1 - \hat \zeta_{ik}}{2 (1 - \hat\gamma_k)},
		\nonumber\\
		&\hat s_{(i,\bm{\nu}_k)}
		=
		\frac{\partial}{\partial \bm{\nu}_k}
		\mathbb{E}[\log h(u_i;\bm{\nu}_k)
		\mid Y_i=y_i,\hat{\bm{\Theta}}_k^{(m)}].
		\label{eq:14}
	\end{align}
	\noindent
	Note the block part of $\hat s_{(i,\bm{\nu}_k)}$ has some integrals which calculated numerically. Standard errors of $\hat{\bm{\Theta}}$ are extracted from the square root of the diagonal elements of the inverse of $\hat{\bm{I}}_0$.

	\section{Simulation studies based on Monte-Carlo}\label{sec.4}
	In this section some Monte-Carlo simulation studies are considered to verify the performances of the MPL estimates. The proposed ECME algorithm has been implemented in the R software (\cite{R2025}) version 4.5.2 with a core i7 760 processor 2.8 GHz.  
	
	In particular, we conduct three simulation experiments using the proposed censored model \eqref{eq:3}, generating censored samples with censoring levels of $p = 10\%, 15\%$, and $30\%$. It is noteworthy that the results across these censoring levels were comparable; therefore, for brevity, we report the results for a representative censoring level in each study.
	
	\medskip
	To explain the mechanism of fulfills the non-informative condition of censored sampling, we consider the following algorithm recommended by \cite{Gomez}.
	
	\begin{remark}\label{rem3}
		Censoring algorithm of the data are considered as following stetps:
		\begin{mdframed}[style=MyFrame]
			\begin{itemize}
				\item[$\bm{Step 1:}$] Let $\bm{y} = (y_1,\dots,y_n)^\top$, be the response observations and let the censoring level $p$, the number of censored samples is $\mathcal{N}= [n \times p]+1$, where $[.]$ denotes the largest integer not greater than, operator. Then, using $sample(\ )$ function in $\mathtt{R}$ software to draw a sample with size of $\mathcal{N}$ from sample sizes $\{1, 2, \cdots, n\}$ without replacement to generate an index censored set $\mathcal{CS}$, 
				\item[$\bm{Step 2:}$] For each $i= 1, \ldots,n$, if $i \in \mathcal{CS}$, generate the $u_i^{(1)} \sim \mathcal{U}(0,c)$ and $u_i^{(2)} \sim \mathcal{U}(0,c)$ independently, where $\mathcal{U}(\alpha,\beta)$ denotes the uniform distribution on interval $(\alpha,\beta)$,   
				\item[$\bm{Step 3:}$] For each $i \in \mathcal{CS}$ set $c_{i1} = \max(y_i - u_i^{(1)}, y_i + u_i^{(2)} -c)$ and $c_{i2} = \min(y_i + u_i^{(2)}, y_i - u_i^{(1)}+c)$, and so $\mathbf{w}_i=(c_{i1},c_{i2})^\top$,
				and for $i \notin \mathcal{CS}$ set $\mathbf{w}_i=y_i$.
			\end{itemize}
		\end{mdframed}
	\end{remark}	
	
	\subsection{Recovery and asymptotic properties of the MPL estimates}\label{sec4.1} 
    To examine the recovery of model's parameters and asymptotic properties of the MPL parameter estimates of the TP-SMN-MoCPLE model obtained via the ECME algorithm, a simulation study  with $K=2$ and following presumed parameters is performed,
    \[
    \bm\beta_1=(\beta_{01}, \beta_{11})^\top=(1, -2)^\top,\;\;\bm\beta_2=(\beta_{02}, \beta_{21})^\top=(2, 3)^\top, \;\; \sigma_1 =1, \;\; \sigma_2 =2,\;\;
    \]
    \[
    \gamma_1 =0.3, \;\; \gamma_2 =0.7,\;\;\bm\tau_1=(0, 0, 0)^\top, \bm\tau_2=(1, 0.2, 2)^\top,
    \]
    $\nu_1=3$ and $\nu_2=4$ for the TP-T-MoCPLE and TP-SL-MoCPLE models, and $(\nu_1,\nu_2)^\top=(0.7,0.7)^\top$ and $(\nu_1,\nu_2)^\top=(0.3,0.3)^\top$ for the TP-CN-MoCPLE model. The non-parametric $g_1(t)=\cos(0.3\pi t)$ and $g_2(t)=\sin(0.16\pi t)$ for $t \in [0,12]$, for the first and second components, receptively.
    We set up $\bm x_i = (1, x_{i1})^\top$ and $\bm r_i = (1, r_{i1}, r_{i2})^\top$, such that $x_{i1} \sim \mathcal{U}(2,4)$, 
    $r_{i1} \sim \mathcal{U}(-2,1)$ and $r_{i2} \sim \mathcal{U}(-1,1)$, for sample sizes $n = 100, 250, 400, 700, 1000$, and then using the Remark \ref{rem3} with $c=1$ and level of censoring $p=15\%$ on the simulated datasets, and the ECME algorithm described in Section \ref{sub.3.2}
    is preformed to carry out the MPL parameter estimates. To investigate the parameters recovery with 1000 replications, we consider the bias (BIAS) and the mean squared error (MSE): 
    \[
    \text{BIAS}(\hat{\theta}_j)=\frac{1}{1000}\sum_{j=1}^{1000}(\hat{\theta}_j - \theta_{real})\qquad
    \text{and} \qquad \text{MSE}(\hat{\theta}_j)=\frac{1}{1000}\sum_{j=1}^{1000}(\hat{\theta}_j - \theta_{real})^2,
    \]
    where $\hat{\theta}_j $ denotes the estimate of a specific model's parameter at the $j$th replication.
    
\begin{figure}\vspace{-1cm}
    	\begin{center}
    		\includegraphics[height=4.7cm, width=16cm]{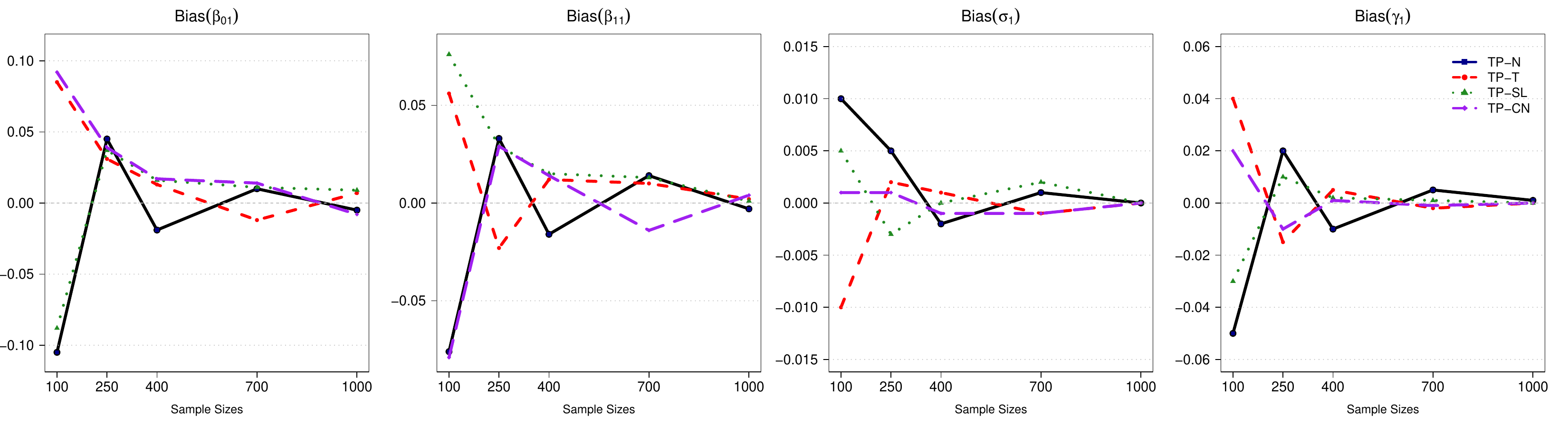}\vspace{-0.1cm}
    		\includegraphics[height=4.7cm, width=16cm]{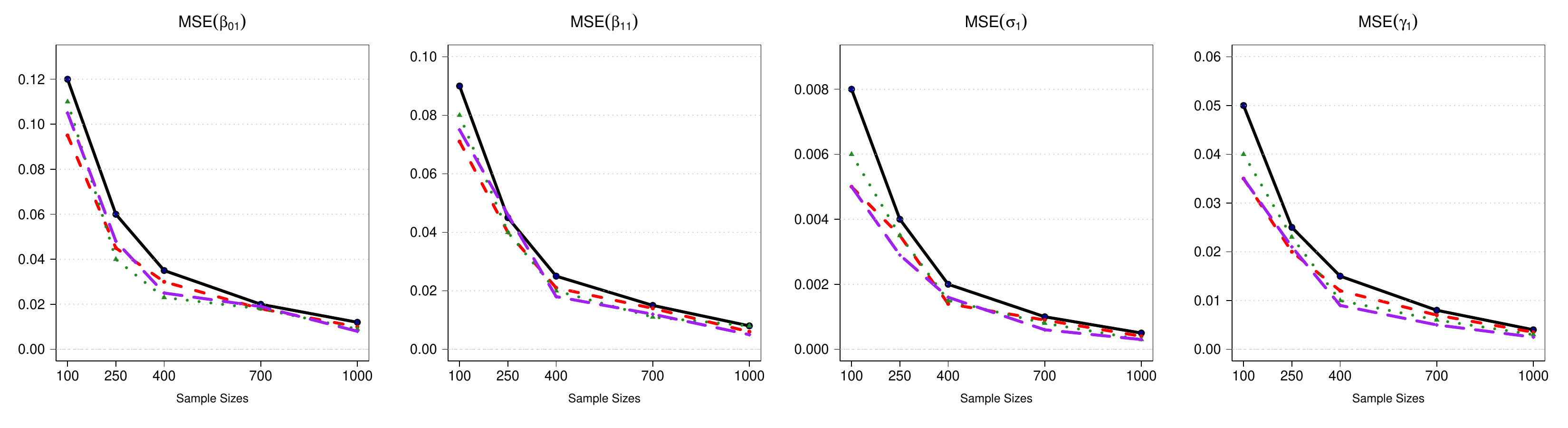}
    		\includegraphics[height=4.7cm, width=16cm]{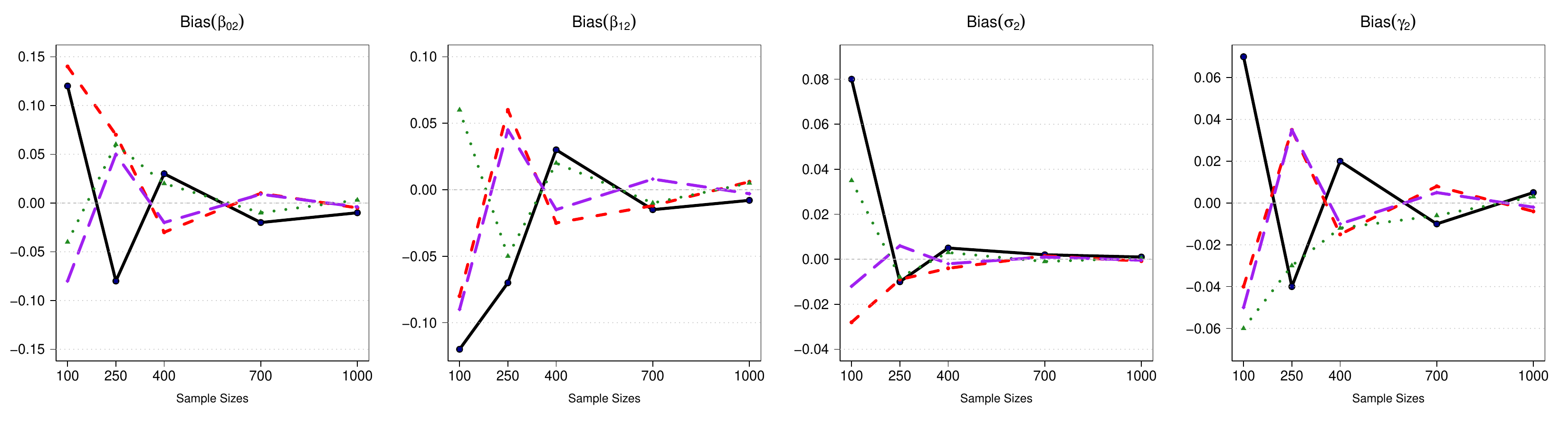}
    		\includegraphics[height=4.7cm, width=16cm]{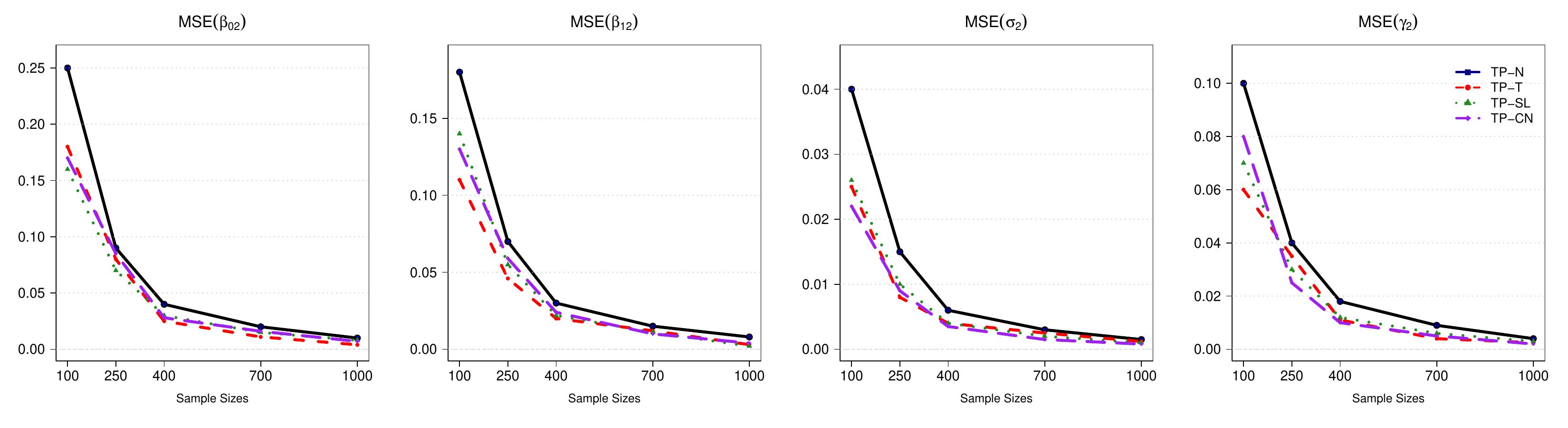}
    	\end{center}
    	\vspace{-0.5cm}
    	\caption{The BIAS and MSE plots of $\bm\beta_k$, $\sigma_k$ and  $\gamma_k$ for $k=1,2$ estimates for the TP-SMN-MoCPLE model with censoring level $15\%$.}\label{fig1}
    \end{figure}

\begin{figure}\vspace{-0.5cm}
    	\begin{center}
    		\includegraphics[height=4.5cm, width=16cm]{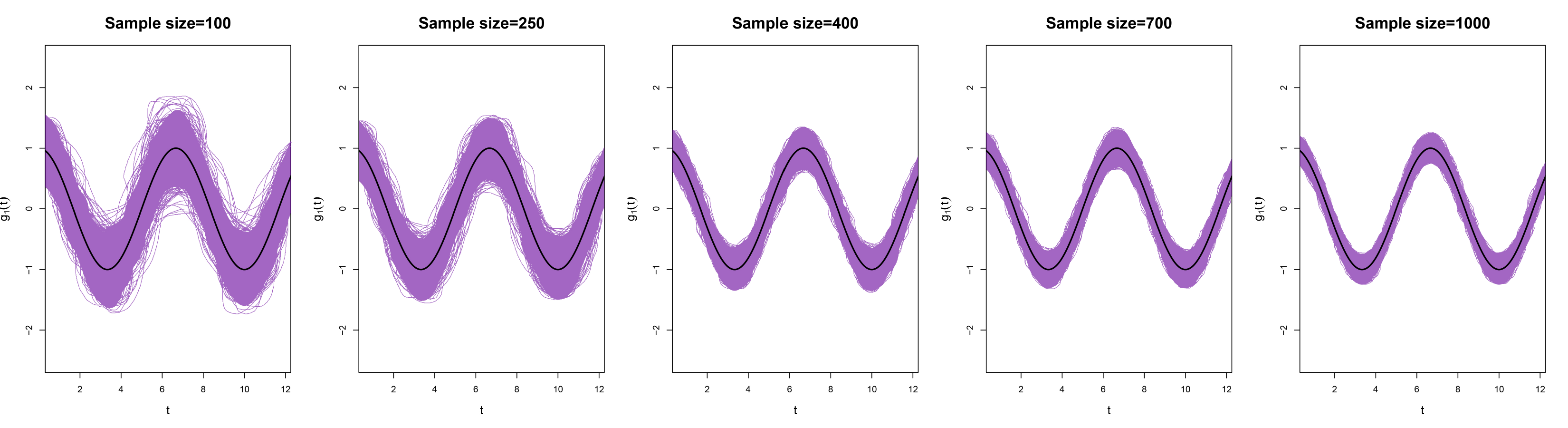}
    		\includegraphics[height=4.5cm, width=16cm]{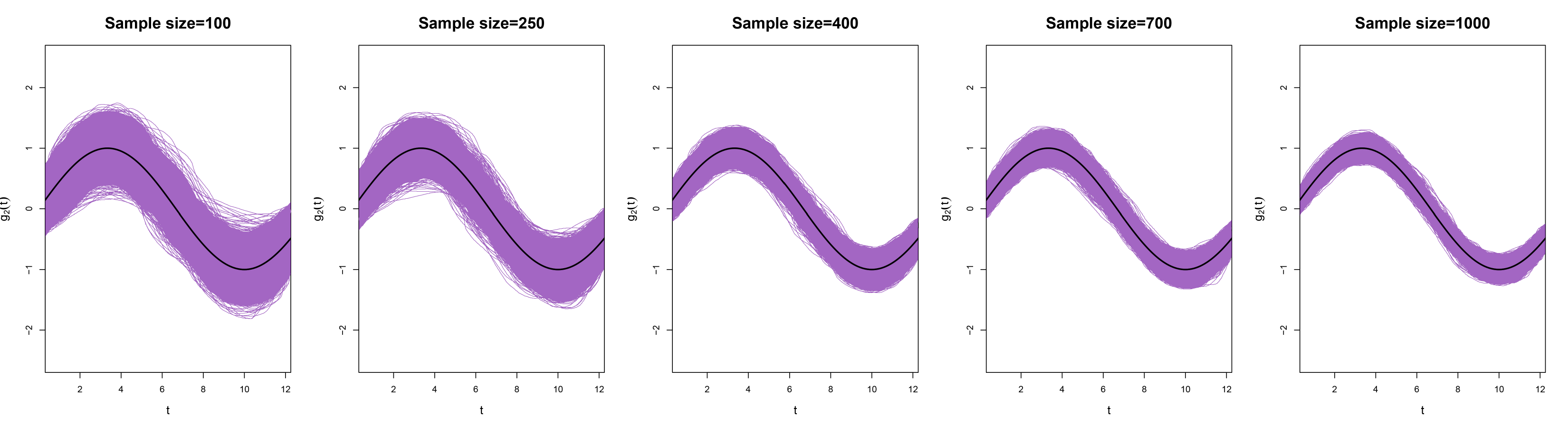}
    	\end{center}
    	\vspace{-0.5cm}
    	\caption{The BIAS and MSE plots of $\bm{g}_k; k=1,2$ estimates for the TP-T-MoCPLE model with censoring level $15\%$.}\label{fig2}
    \end{figure}

    \medskip
    Figure \ref{fig1} displays the BIAS and MSE plots of the MPL estimates of the model's parameters $\bm\beta_k$, $\sigma_k$ and  $\gamma_k$ for $k=1,2$, of the TP-N-MoCPLE, TP-T-MoCPLE, TP-SL-MoCPLE and TP-CN-MoCPLE models for the censored datasets. It can be observed that the MPL estimates of the parameters have very small (around zero) BIAS for all sample sizes, and, as $n$ increases the MSE of MPL estimates tend to zero. In the case of TP-T-MoCPLE model, the plots in Figure \ref{fig2} also shows the descending trends and the variability of the nonparametric function estimates increases, and  diminishes with increasing sample size. Similar results were obtained for the non-parametric parts of the other TP–SMN–CPLMs.These behaviours indicate the empirically consistency of the
    our MPL estimates of the model parameters via the ECME algorithm. 

	\subsection{Model selection performance in the presence of outliers and asymmetry}\label{sec4.2}
	Optimal number of experts $K$ in the MoE models is one the main challenges. In this simulation study we generate 1000 sample of size $n=300$ from the MoCPLE model that error terms on each expert follow the skew-\textit{t} (SMSN; \cite{Branco}), which is a skew and heavy-tailed distribution with similar properties of the TP-T distribution in the class of the TP-SMN family. We call the proposed model as ST-MoCPLE model, which same as some members of our models, cover the censorious, asymmetry and outliers. But note that ST-MoCPLE model has not conditional expectations and so the MPL estimates are not exist, but we expect our MoCPLE models could detect the true number of experts.
	
\begin{figure}\vspace{-0.5cm}
		\begin{center}
			\includegraphics[height=7.2cm, width=16cm]{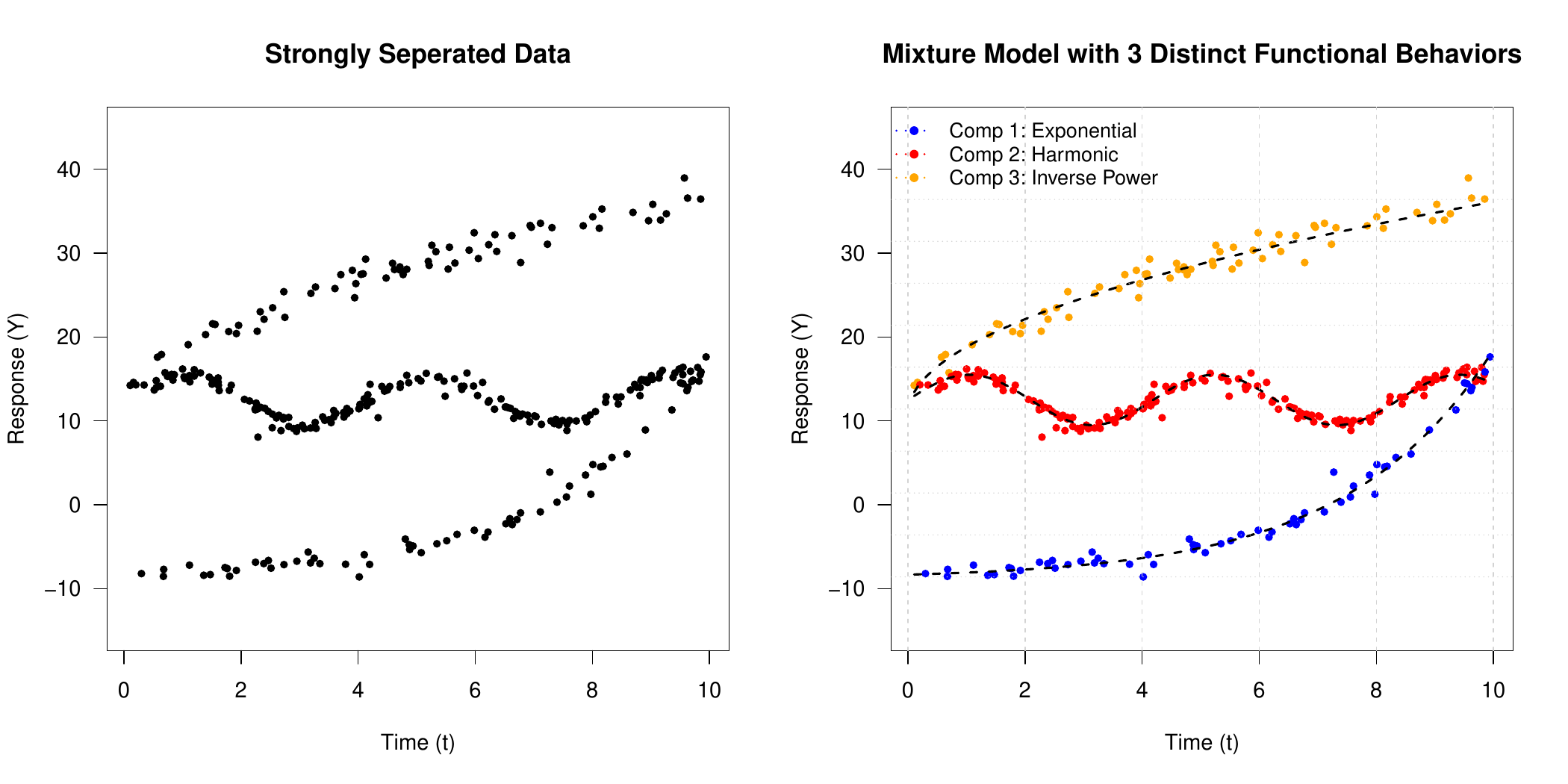}
			\includegraphics[height=7.2cm, width=16cm]{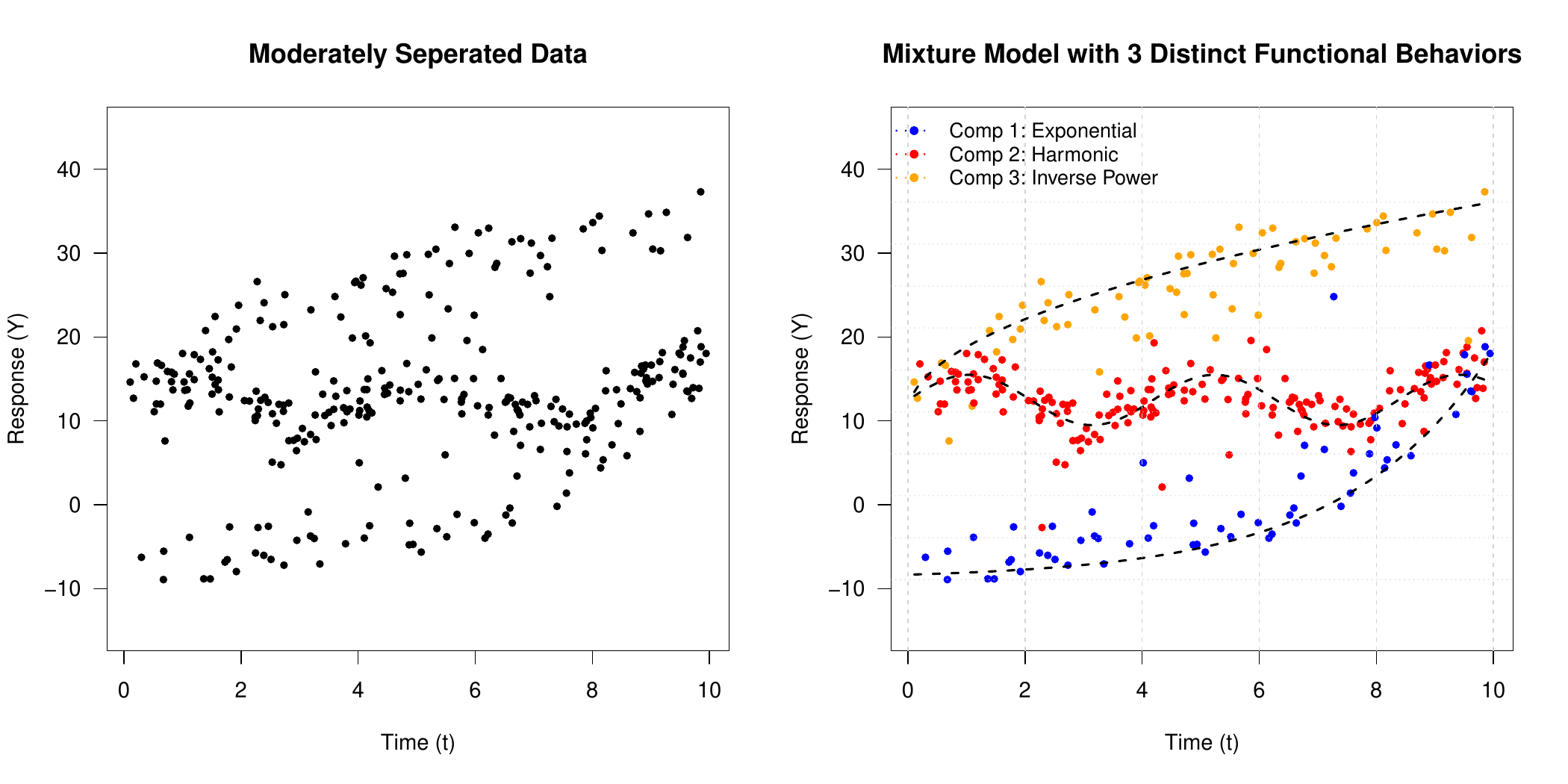}
			\includegraphics[height=7.2cm, width=16cm]{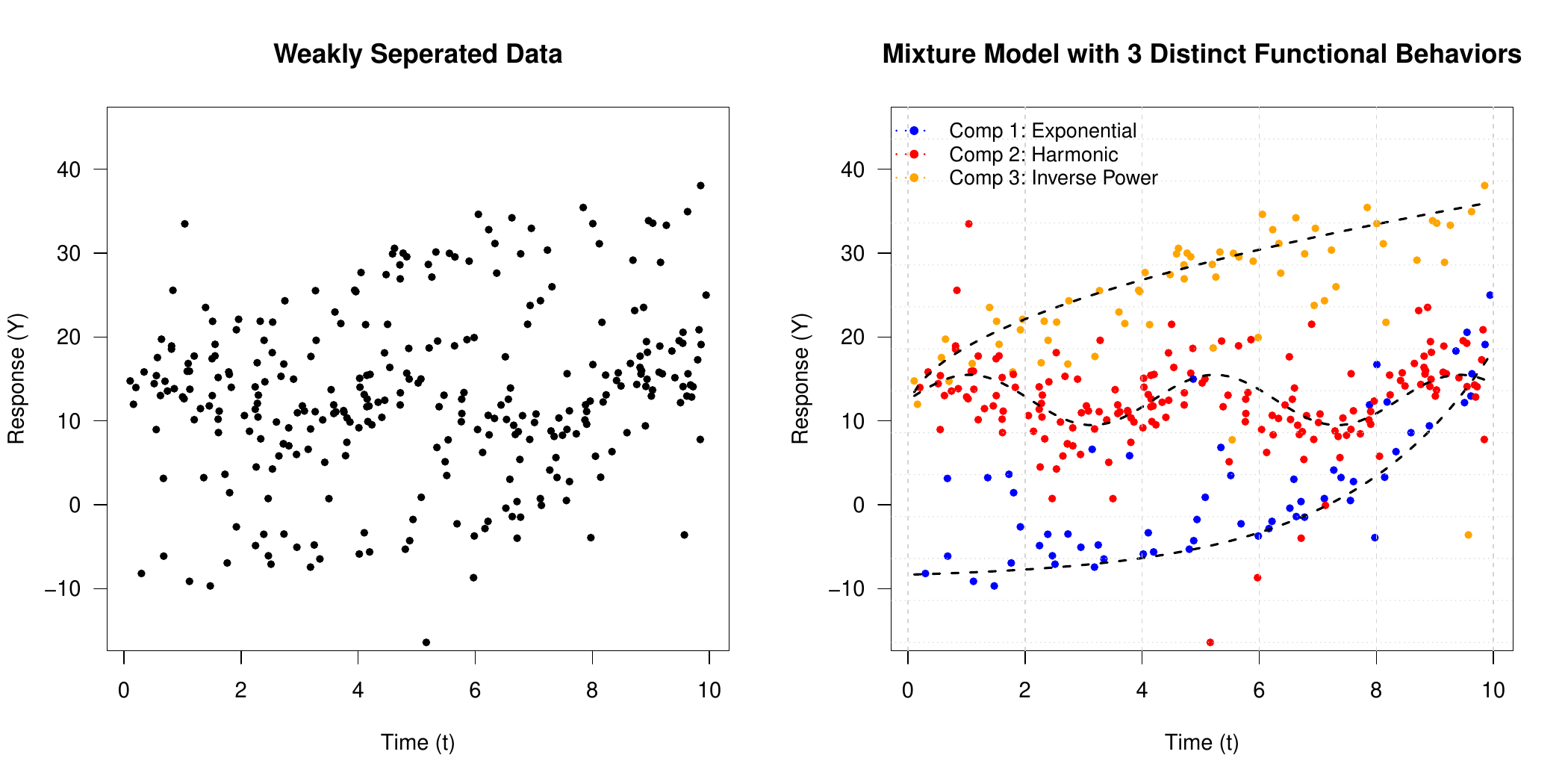}
		\end{center}
		\vspace{-0.5cm}
		\caption{The BIAS and MSE plots of $\bm{g}_k; k=1,2,3$ estimates for the TP-SMN-MoCPLE model with censoring level $15\%$.}\label{fig3}
	\end{figure}

	We consider ST-MoCPLE model with three number of experts $K=3$ and following presumed parameters:
	
	\[
	\;\;\bm\tau_1=(0, 0)^\top, \bm\tau_2=(1, -2)^\top, \bm\tau_3=(1, 2)^\top,\bm\beta_1=(-10, 0.5)^\top,\;\;\bm\beta_2=(12, 0.1)^\top,\;\;\bm\beta_3=(17, -0.8)^\top,
	\]
	with exponential, harmonic and inverse power non-parametric parts of experts for the first, second and third experts given by:
	\[
	g_1(t)=0.5\exp(0.4t), \;\; g_2(t)=3\sin(1.5t), \;\; g_3(t)=8t^\frac{1}{2}, 
	\] 
	for $t \in [0,10]$, $\bm x_i = (1, x_{i1})^\top$ and $\bm r_i = (1, r_{i1})^\top$, such that $x_{i1} \sim \mathcal{U}(2,4)$, 
	$r_{i1} \sim \mathcal{U}(1,2)$, with $\nu_1=\nu_2=\nu_3=4$ and other remaining parameters are chose such that have three following scenarios: 
	
	\begin{itemize}
		\item {Strongly separated experts (S.S.E.):} 
		\[
		\;\; \sigma_1 =\sigma_2 =\sigma_3 =1,\;\; \alpha_1=-1, \;\; \alpha_2=0, \;\; \alpha_3=+1,
		\]
		\item {Moderately separated experts (M.S.E.):} 
		\[
		\;\; \sigma_1 =\sigma_2 =\sigma_3 =3,\;\; \alpha_1=-2, \;\; \alpha_2=0, \;\; \alpha_3=+2,
		\]
		\item {Weakly separated experts (W.S.E.):} 
		\[
		\;\; \sigma_1 =\sigma_2 =\sigma_3 =5,\;\; \alpha_1=-3, \;\; \alpha_2=0, \;\;    \alpha_3=+3,
		\]
	\end{itemize}  
	where $\alpha_k, k=1,2,3$ are the skewness parameters of skew-\textit{t} distribution on each expert (Example of generated samples of the above three scenarios are shown in Figure \ref{fig3}), and then using the Remark \ref{rem3} with $c=1$ and level of censoring $p=10\%$ on the simulated datasets.
	
	In the following analyses, the number of mixture components (experts), $K$, is assumed to be unknown. Accordingly, we fitted the TP-N-MoCPLE, TP-T-MoCPLE, TP-SL-MoCPLE, and TP-CN-MoCPLE models to the simulated data. This fitting was performed for each model assuming $K=1, 2, 3, 4,$ and $5$ experts in each replication. The results obtained from the AIC criterion, which indicate the percentage of correct specification for the number of experts among the different replications, are presented in Table \ref{tab11}. As shown in Table \ref{tab11}, the TP-T-MoCPLE, TP-SL-MoCPLE, and TP-CN-MoCPLE models outperform the TP-N-MoCPLE model in determining the number of components, likely due to the heavy-tailed nature of the generated data. Notably, the TP-T-MoCPLE model achieves a better fit than the other models, particularly when the optimal number of experts is three ($K=3$).
	
	\begin{table*}[!t]
		\centering{\caption{Simulation results, based on 1000 replications, for performance comparison of the TP-SMN-MoCPLE sub-models
				to the generated data from the MoE-SGIG-CR model.\label{tab11}}
			\scalebox{0.64}{\hspace{-1cm}
				\begin{tabular}{lccccccccccccccccccc}
					\hline
					Separating Level	& Model  & $K=1$ && $K=2$ && $K=3$ && $K=4$ && $K=5$\\
					\hline
					S.S.E.	& TP-N-MoCPLE & \quad 2.3\% && \quad 4.1\% && \quad 2.7\% && \quad 1.2\% 
					&& \quad 1.8\% \\
					& TP-T-MoCPLE & 67.2\% && 60.4\% && $\bm{67.3\%}$ && 58.4\% && 47.9\% \\
					& TP-SL-MoCPLE & 20.3\% && 20.5\% && 21.8\% && 18.9\% && 27.6\% \\
					& TP-CN-MoCPLE & 10.2\% && 15.0\% && \quad 8.2\% && 21.5\% && 22.7\% \\
					\\
					M.S.E.	& TP-N-MoCPLE & \quad 4.3\% && \quad 3.8\% && \quad 1.9\% && \quad 1.1\% && \quad 0.9\% \\
					& TP-T-MoCPLE & 58.7\% && 59.2\% && $\bm{62.8\%}$ && 39.1\% && 40.6\% \\
					& TP-SL-MoCPLE & 24.1\% && 20.5\% && 12.4\% && 22.6\% && 20.8\% \\
					& TP-CN-MoCPLE & 12.9\% && 16.5\% && 22.9\% && 37.2\% && 37.7\% \\
					\\
					W.S.E.	& TP-N-MoCPLE & \quad 1.8\% && \quad 2.3\% && \quad 1.5\% && \quad 2.1\% 
					&& \quad 1.9\% \\
					& TP-T-MoCPLE & 34.1\% && 30.8\% && $\bm{69.3}\%$ && 34.1\% && 39.2\% \\
					& TP-SL-MoCPLE & 28.4\% && 30.1\% && 30.6\% && 33.3\% && 30.1\% \\
					& TP-CN-MoCPLE & 35.7\% && 37.1\% && \quad 5.2\% && 30.5\% && 28.8\% \\	   
					\hline
				\end{tabular}
		}}
	\end{table*}

\section{Real datasets analysis}\label{sec.5}
In this section to asses the performance the proposed model and estimates we consider two real applications. The first dataset is New York Air Quality Measurements, well known as \texttt{airquality} which is meteorological data and available in an R package \texttt{datasets}, and second is Labor Force Participation Data, well-known as \texttt{PSID1976} dataset which is economical data, and available in an R package \texttt{AER}. 
\subsection{New York Air Quality Measurements Data}\label{sec5.1}
For the first empirical analysis, we utilized the well-known \texttt{airquality} dataset, which contains daily air quality measurements recorded in New York City from May 1st to September 30th, 1973. This dataset, originally provided by the New York State Department of Conservation (for ozone data) and the National Weather Service, consists of a data frame with 153 observations (\cite{Chambers}). After removing missing values of auxiliaries and gating variables, our analysis is conducted on 111 complete observations. We consider the following variables:

\begin{itemize}
	\item Response variable (\textbf{Ozone}; $y$): Mean ozone concentration in parts per billion (ppb), measured between 1300 and 1500 hours at Roosevelt Island.
	\item First auxiliary variable (\textbf{Solar.R}; $x_1$): Solar radiation in Langleys, measured in the frequency band of 4000–7700 Angstroms between 0800 and 1200 hours at Central Park.
	\item Second auxiliary variable (\textbf{Month}; $x_2$): The month of observation (May to September), included as a categorical covariate in the linear component.
	\item First gating variable (\textbf{Wind}; $r_1$): Average wind speed in miles per hour (mph), recorded at 0700 and 1000 hours at LaGuardia Airport.
	\item Functional variable (\textbf{Temp}; $t$): Maximum daily temperature in degrees Fahrenheit, recorded at LaGuardia Airport, used to construct the nonparametric basis matrix.
\end{itemize}
\begin{figure}\vspace{-0.5cm}
	\begin{center}
		\includegraphics[height=18cm, width=16.5cm]{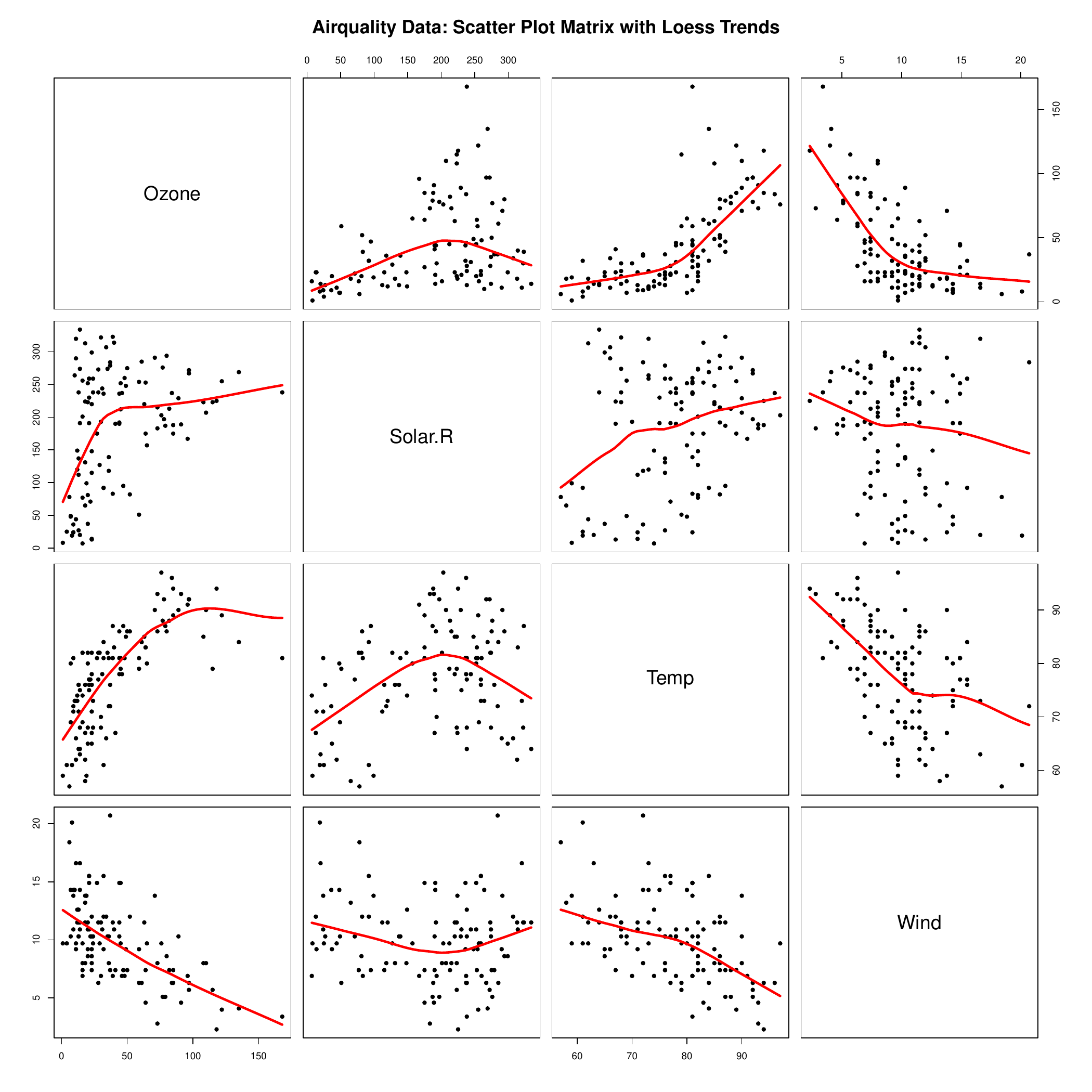}
	\end{center}
	\vspace{-0.5cm}
	\caption{Pairwise scatterplots of Ozone, Solar.R, Temp, and Wind with (red) smooth lines illustrating the general trend between variables.}\label{fig4}
\end{figure}
    \noindent
    The trends observed in the pairwise plots in Figure \ref{fig4}, visualized via nonparametric regression curves (LOESS smoothers), indicate that a purely linear assumption may be inadequate to capture the complex relationships, particularly the nonlinear behavior of temperature and wind speed. To address this, we consider our semi-parametric framework. Specifically, we define the linear component using Solar.R and Month, while the nonlinear effects are modeled through the functional variable (Temp) and the gating variable (Wind). In the next step, we apply the TP-SMN-MoCPLE models with several experts to estimate the optimal model structure under an interval-censoring scheme.

In our modelling approach, we employed auxiliary regression variables comprising \texttt{Solar.R} and \texttt{Month} to define the linear component of the model. Furthermore, the gating expert variables selected for the model were \texttt{Wind} and the functional basis derived from \texttt{Temp}. These specific variables were chosen due to their established influence on air quality dynamics and their frequent use in related atmospheric studies. This dataset, along with the selected variables, provides a comprehensive basis for investigating the relationships between meteorological factors and air pollution levels.

Using the procedure described in Remark \ref{rem3} with c=1 and an interval censoring level of 10\% (including 20\% of the censored observations assigned to wider intervals to simulate potential outliers), we fitted the TP-SMN-MoCPLE models to the data, considering components ranging from $K=1$ to $K=4$. Based on the AIC criterion, the two-expert ($K=2$) model was selected as the optimal fit. To the best of our knowledge, Mixture of Experts models based on semi-parametric frameworks and non-normal distributions have not yet been introduced in the literature, precluding a direct comparative analysis with existing studies. However, the incorporation of both skew and symmetric distributions within our proposed model provides a comprehensive representation of various data behaviors, thereby ensuring the model's flexibility and robustness.

\begin{table}
	\centering{\caption{MPL estimates with corresponding approximate standard errors(SE) together with their AIC and BIC values.\label{real1}}
		\scalebox{0.7}{\hspace{-0.5cm}
			\begin{tabular}{ccccccccccccccccc}
				\hline
				&& \multicolumn{2}{c}{TP-N-MoCPLE} &&
				\multicolumn{2}{c}{TP-T-MoCPLE} & & \multicolumn{2}{c}{TP-SL-MoCPLE} & & \multicolumn{2}{c}{TP-CN-MoCPLE}\\
				\cline{3-4}\cline{6-7}\cline{9-10}\cline{12-13}
				Parameter $\downarrow$ && Estimates & SE && Estimates & SE && Estimates & SE && Estimates & SE\\
				\hline 
				$\beta_{10}$ && 14.8654 & 0.7343 && 12.2424 & 0.6104 && 12.8632 & 0.7014 && 13.0142 & 0.9132\\
				$\beta_{11}$ && 0.1201 & 0.0432 && 0.0480 & 0.0065 && 0.0491 & 0.0036 && 0.0501 & 0.0044\\			  
				$\beta_{12}$ && 0.1067 & 0.0945 && 0.5709 & 0.0213 && 0.5811 & 0.0196 && 0.6002 & 0.0201 \\
				$\beta_{20}$ && -1.0533 & 0.1456 && -0.6205 & 0.0229 && -0.6511 & 0.0184 && -0.6801 & 0.0252\\
				$\beta_{21}$ && 0.4711 & 0.0565 && 0.2201 & 0.0321 && 0.2196 & 0.0252 && 0.2297 & 0.0364\\
				$\beta_{22}$ && 7.1028 & 0.9231 && 5.6859 & 0.8524 && 5.7532 & 0.9432 && 5.7604 & 0.8603\\
				$\sigma_{1}$ && 20.2011 & 1.9309 && 16.2606 & 1.0763 && 16.7109 & 1.1201 && 16.7301 & 1.1172\\
				$\sigma_{2}$ && 39.7212 & 3.3028 && 34.8594 & 2.2010 && 35.0172 & 2.2654 && 35.2191 & 2.4021\\
				$\nu_1$ && -- & -- && 14.7762 & 0.7036 && 14.5673 & 0.8001 && 0.8389 & 0.0021\\
				$\nu_2$ && -- & -- && 4.0576 & 0.1453 && 4.2011 & 0.1351 && 0.9001 & 0.0073\\
				$\delta_1$ && -- & -- && -- & -- && -- & -- && 0.8243 & 0.0312 \\
				$\delta_2$ && -- & -- &&  -- & -- && -- & -- && 0.8637 & 0.0211\\
				$\gamma_1$ &&  0.4548 & 0.0076  && 0.4985 & 0.0024 && 0.4891 & 0.0032 && 0.4893 & 0.0031 \\
				$\gamma_2$ && 0.3683 & 0.0089  && 0.4759 & 0.0029 && 0.4702 & 0.0037 && 0.4731 & 0.0040 \\
				$\tau_{20}$ && 6.8746 & 2.2191 && 5.2849 & 0.6222 && 5.1873 & 0.4563 && 5.5631 & 0.6101\\
				$\tau_{21}$ && -1.263 & 0.0431 && -0.7127 & 0.0283 && -0.7990 & 0.0131 && -0.7488 & 0.0236\\
				\hline
				AIC && 1119.203 &&& 1055.603 &&& \textbf{1053.803} &&& 1061.203 \\
				BIC && 1314.289 &&& 1258.818 &&& \textbf{1257.018} &&& 1267.127 \\
				RI  && 0.5123   &&&  0.5214  &&&  \textbf{0.5323}  &&& 0.5118 \\ 
				JCI && 0.3676   &&&  0.3847  &&&  0.4029  &&& 0.3713 \\	 
				\hline
			\end{tabular}   
	}}
\end{table}

\begin{figure}[htbp]
	\centering
	\includegraphics[height=7cm, width=7cm]{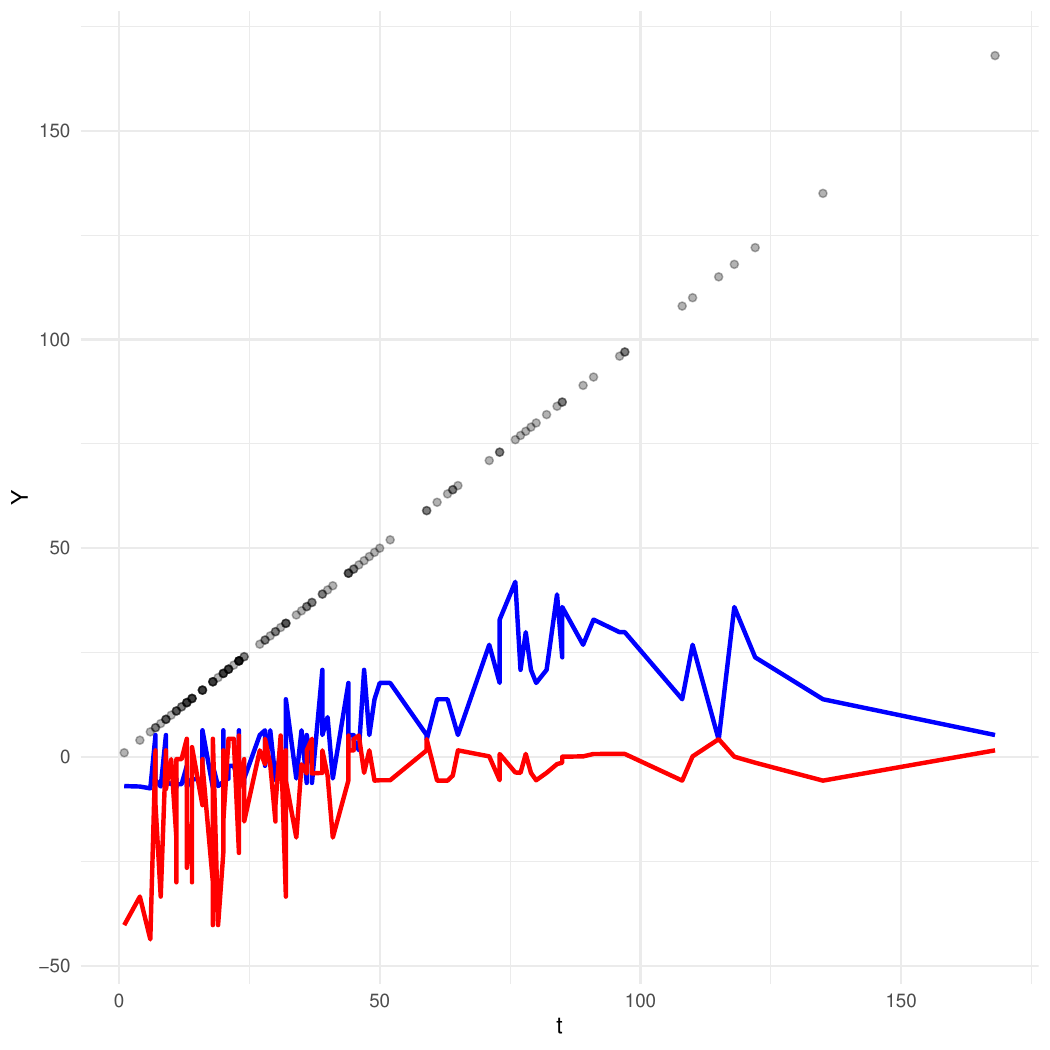} 
	\includegraphics[height=7cm, width=7cm]{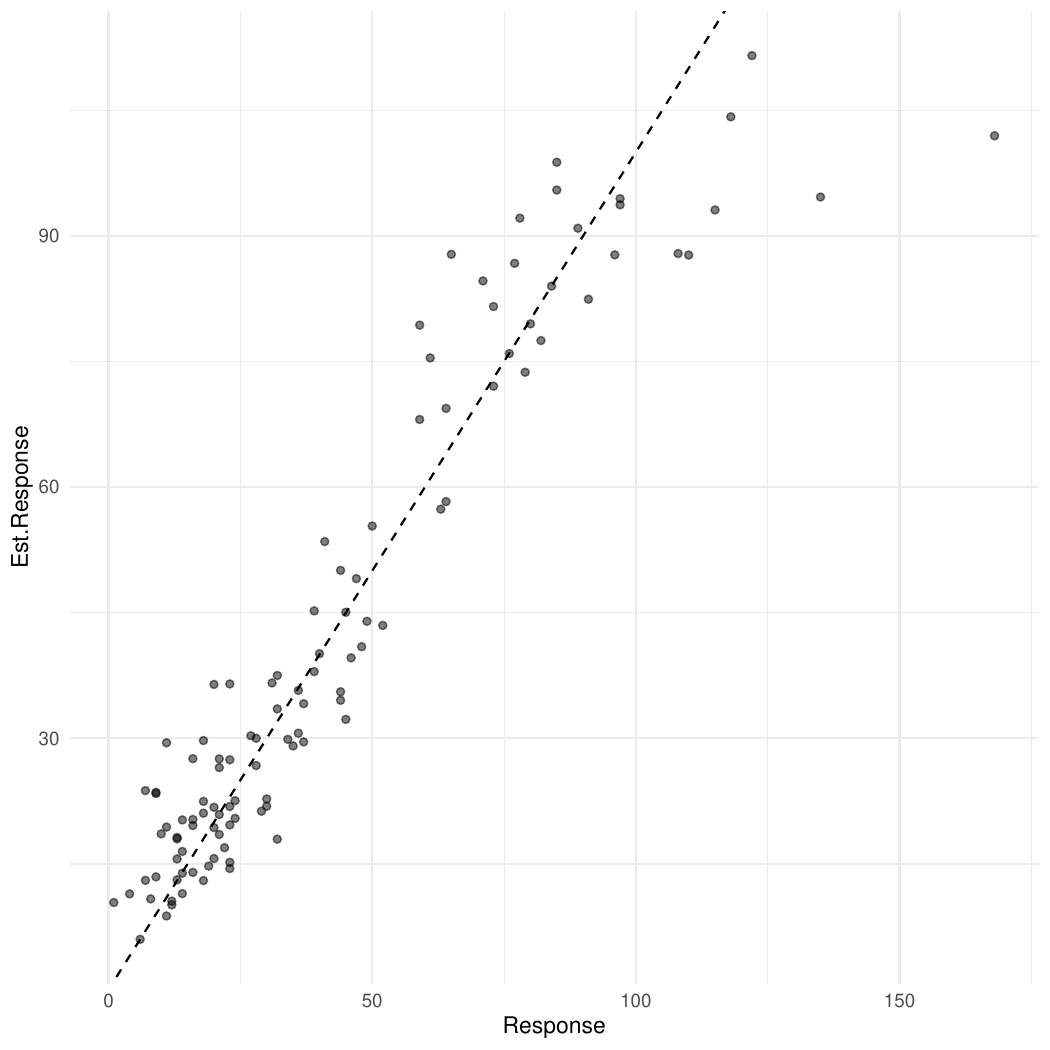}
	\vspace{0.2cm} 
	\caption{Model fitting diagnostics. Left: Non-parametric functions $\text{g}_k, k=1,2$, for the two experts. Right: Scatter plot of observed versus fitted responses.}
	\label{fig5}
\end{figure}

Table \ref{real1} presents the MPL estimates for the four considered models. The penalty coefficients $(\hat \lambda_1,\hat \lambda_2)$ for both experts were set to 0.01 and 0.02, using the $\lambda$-grid approach. The substantial disparity between the intercepts points to different underlying distributions for the two experts. Both solar radiation (\texttt{Solar.R}) and month of observation (\texttt{Month}) have positive influence on the mean ozone in both experts; however, the influence of solar radiation is more strong than that of month of observation, and influence of auxiliaries are more effective in the second expert. Regarding the shape parameters, the gamma estimates reveal slight skewness (near symmetry) for the both experts. Additionally, the small values of degrees of freedom on each experts (especially on the second) characterized by heavy tails and the presence of outliers. The coefficients in the gating function also demonstrate that average wind speed (\texttt{Wind}) has positive influence on expert selection. Finally, based on the AIC and BIC (\cite{Schwarz}) values recorded in the Table \ref{real2}, the TP-T-MoCPLE and TP-SL-MoCPLE models demonstrate a superior fit compared to the other models.
Figure \ref{fig5} (left) illustrates the non-parametric functions $\text{g}_k, k=1,2$ for both experts, plotted against the response observations, and Figure \ref{fig5} (right) displays the observed response values plotted against the estimated fitted values. The close alignment of the points along a straight line on Figure \ref{fig5} (right), indicates that the model provides a good fit to the data.

\subsection{Labor Force Participation Data}\label{sec5.2}
To further assess the applicability of our proposed methods, we employ the \texttt{PSID1976} dataset, a widely used cross-section dataset originating from the 1976 Panel Study of Income Dynamics (PSID). This dataset focuses on labor force participation and related socio-economic factors for individuals in 1975. It comprises 753 observations offering a rich context for analyzing factors influencing economic decisions, that for our analysis we choose following variables:

\begin{itemize}
	\item Response variable (\textbf{hwage}; $y$): Husband's average hourly wage in 1975 dollars. This is often used as the dependent variable in seminal studies using this dataset.
	\item First auxiliary variable (\textbf{hhours}; $x_1$): Husband's hours of work in 1975.
	\item Second auxiliary variable (\textbf{heducation}; $x_2$): Husband's education in years.
	\item Second auxiliary variable (\textbf{hage}; $x_3$): Husband's age in years.
	\item First gating variable (\textbf{fincome}; $r_1$): Family income in 1975 dollars, used for constructing property income.
	\item Second gating variable (\textbf{unemp}; $r_2$): Unemployment rate in county of residence, in percentage points. (This is taken from bracketed ranges.)
	\item Functional variable (\textbf{experience}; $t$): Actual years of wife's previous labor market experience.
\end{itemize}
\begin{figure}\vspace{-0.5cm}
	\begin{center}
		\includegraphics[height=18cm, width=17cm]{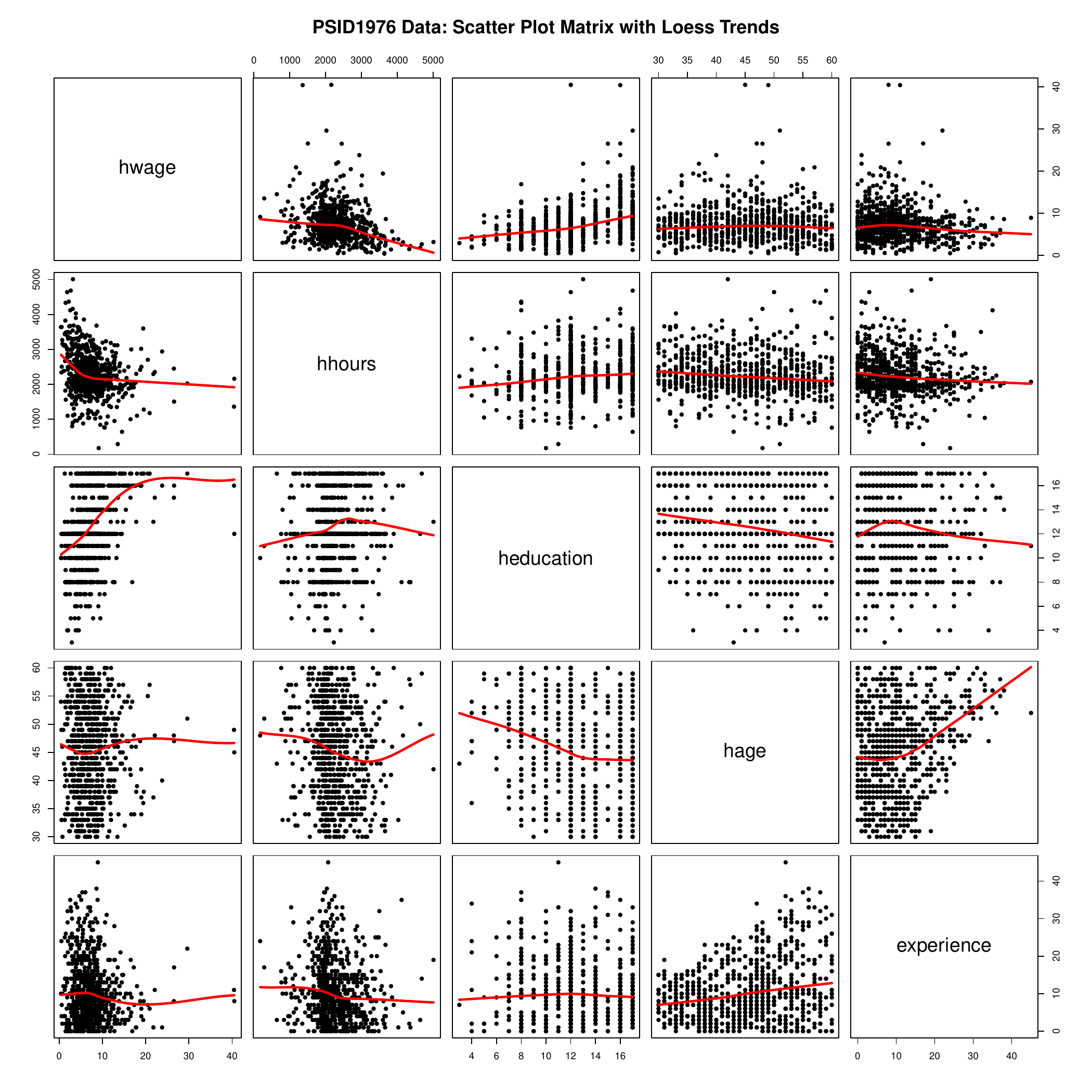}
	\end{center}
	\vspace{-0.5cm}
	\caption{Pairwise scatterplots of hwage, hhours, heducation, hage and experience with (red) smooth lines illustrating the general trend between variables.}\label{fig6}
\end{figure}

This dataset, also known as the "wage rates" of 753 married women dataset \cite{Mroz,  Karlsson}, is particularly valuable for analyzing labor supply decisions. The trends observed in the pairwise plots in Figure \ref{fig6}, visualized via nonparametric regression curves (LOESS smoothers), indicate that a purely linear assumption may be inadequate to capture the complex relationships between the observed variables. We assume the husband's average hourly wage depends on their hours of work, education, age and experience, and using the Remark \ref{rem3} with c=1 and an interval censoring level of 10\% (including 20\% of the censored observations assigned to wider intervals to simulate potential outliers). Therefore, we fit our proposed TP-SMN-MoCPLE models for several possible number of experts. Optimal number of experts $K=3$ were obtained by AIC criteria between possible values $K=1,2,3,4$. Also using the $\lambda$-grid approach, the penalty coefficients $(\hat \lambda_1,\hat \lambda_2,\hat \lambda_3)$ were set to (0.05,0.05,0.05), and MPL estimates of proposed models are provided in the Table \ref{real2}.

\begin{figure}[htbp]
	\centering
	\includegraphics[height=7cm, width=7cm]{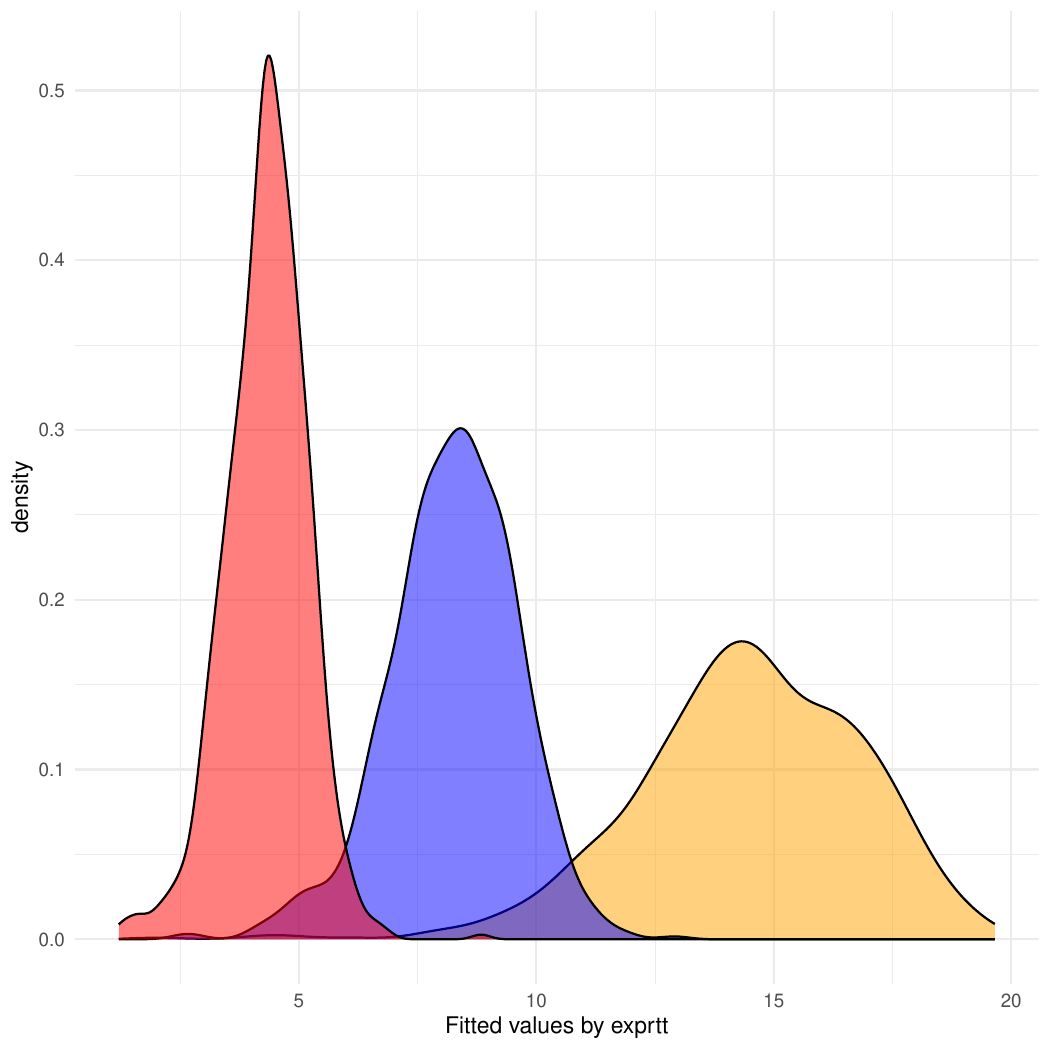} 
	\includegraphics[height=7cm, width=7cm]{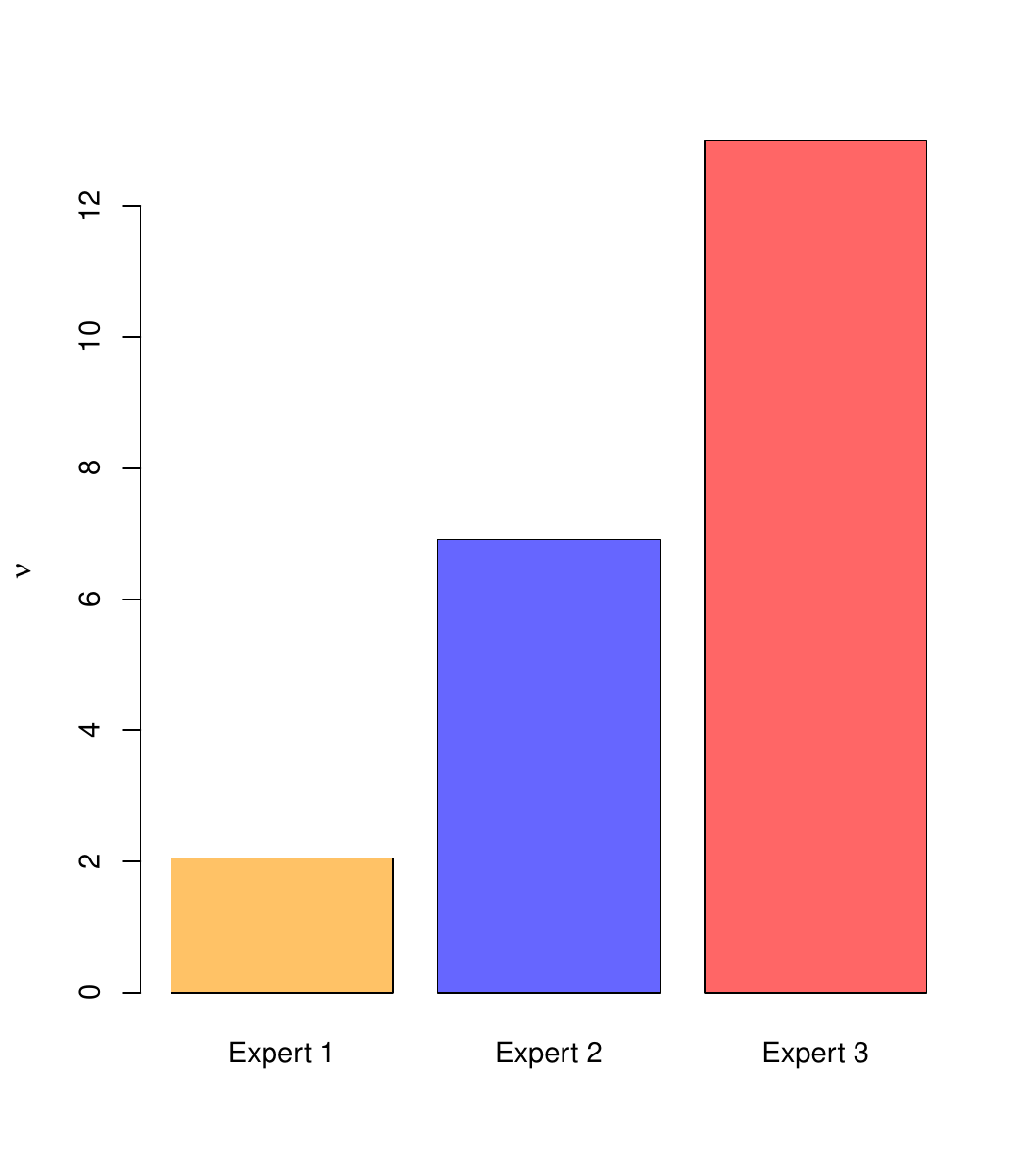}
	\vspace{0.2cm} 
	\caption{Three experts. Left: Non-parametric density estimates of the fitted values across the three experts. Right: Barplot of degrees of freedom on experts.}
	\label{fig7}
\end{figure}

\begin{table}
	\centering{\caption{MPL estimates with corresponding approximate standard errors(SE) together with their AIC and BIC values.\label{real2}}
		\scalebox{0.7}{\hspace{-0.5cm}
			\begin{tabular}{ccccccccccccccccc}
				\hline
				&& \multicolumn{2}{c}{TP-N-MoCPLE} &&
				\multicolumn{2}{c}{TP-T-MoCPLE} & & \multicolumn{2}{c}{TP-SL-MoCPLE} & & \multicolumn{2}{c}{TP-CN-MoCPLE}\\
				\cline{3-4}\cline{6-7}\cline{9-10}\cline{12-13}
				Parameter $\downarrow$ && Estimates & SE && Estimates & SE && Estimates & SE && Estimates & SE\\
				\hline 
				$\beta_{10}$ && 15.3431 & 0.9873 && 13.8067 & 0.8112 && 13.4463 & 0.9038 && 13.9983 & 0.9432\\
				$\beta_{11}$ && -0.0412 & 0.0023 && -0.0012 & 0.0003 && -0.0014 & 0.0005 && -0.0013 & 0.0003\\			  
				$\beta_{12}$ && 0.9082 & 0.0887 && 0.5812 & 0.0232 && 0.5814 & 0.0306 && 0.5902 & 0.0311 \\
				$\beta_{13}$ && 0.1067 & 0.0945 && -0.0602 & 0.0054 && -0.0644 & 0.0065 && -0.0612 & 0.0069 \\
				$\beta_{20}$ && 9.9109 & 0.8575 && 8.8633 & 0.2429 && 8.9018 & 0.2871 && -0.6801 & 0.0252 \\
				$\beta_{21}$ && 0.4711 & 0.0565 && -0.0019 & 0.0004 && -0.0021 & 0.0004 && -0.0019 & 0.0005\\
				$\beta_{22}$ && 7.1028 & 0.9231 && 0.2907 & 0.0231 && 0.3010 & 0.0306 && 0.2958 & 0.0149\\
				$\beta_{23}$ && 7.1028 & 0.9231 && 0.0028 & 0.0006 && 0.0030 & 0.0007 && 0.0029 & 0.0007\\
				$\beta_{30}$ && 7.3918 & 0.9001 && 5.1371 & 0.6739 && 5.2101 & 0.6881 && 5.3010 & 0.7781 \\
				$\beta_{31}$ && -0.0722 & 0.0146 && -0.0011 & 0.0003 && -0.0012 & 0.0005 && -0.0014 & 0.0006\\
				$\beta_{32}$ && 0.5093 & 0.0882 && 0.1690 & 0.0332 && 0.1705 & 0.0367 && 0.1686 & 0.0300\\
				$\beta_{33}$ && -0.0542 & 0.0655 && -0.0080 & 0.0005 && -0.0082 & 0.0004 && -0.0089 & 0.0004\\
				$\sigma_{1}$ && 18.3029 & 1.0817 && 5.4143 & 0.5387 && 5.4396 & 0.5611 && 5.4146 & 0.6048\\
				$\sigma_{2}$ && 6.6355 & 0.9983 && 2.3106 & 0.1872 && 2.4018 & 0.1998 && 2.2827 & 0.2018\\
				$\sigma_{3}$ && 7.0948 & 1.0008 && 2.7178 & 0.1099 && 2.8018 & 0.1222 && 2.6902 & 0.1100\\
				$\nu_1$ && -- & -- && 2.0500 & 0.0653 && 2.2201 & 0.0809 && 0.0162 & 0.0043\\
				$\nu_2$ && -- & -- && 6.9094 & 0.1908 && 7.2011 & 0.2019 && 0.0210 & 0.0088\\
				$\nu_3$ && -- & -- && 12.9973 & 0.2201 && 13.2031 & 0.2353 && 0.3001 & 0.0104\\
				$\delta_1$ && -- & -- && -- & -- && -- & -- && 0.0621 & 0.0041 \\
				$\delta_2$ && -- & -- &&  -- & -- && -- & -- && 0.1031 & 0.0113\\
				$\delta_3$ && -- & -- &&  -- & -- && -- & -- && 0.2109 & 0.0344\\
				$\gamma_1$ &&  0.3309 & 0.0276  && 0.2738 & 0.0108 && 0.2409 & 0.0099 && 0.2693 & 0.0110 \\
				$\gamma_2$ && 0.3209 & 0.0201 && 0.4133 & 0.0138 && 0.4042 & 0.0221 && 0.3939 & 0.0202 \\
				$\gamma_3$ && 0.7033 & 0.0332 && 0.6267 & 0.0200 && 0.6301 & 0.0248 && 0.6132 & 0.0313 \\
				$\tau_{20}$ && 6.0983 & 1.4533 && 7.1261 & 0.5984 && 7.8736 & 0.7001 && 7.5847 & 0.5666\\
				$\tau_{21}$ && -0.8981 & 0.0431 && -0.0001 & 0.0000 && -0.0002 & 0.0000 && -0.0001 & 0.0000\\
				$\tau_{22}$ && -1.263 & 0.0431 && -0.1217 & 0.0109 && -0.1345 & 0.0145 && -0.1298 & 0.0121\\
				$\tau_{30}$ && 20.1981 & 1.8716 && 14.7412 & 0.9817 && 15.0911 & 1.0081 && 15.0918 & 0.9987\\
				$\tau_{31}$ && -0.8711 & 0.0601 && -0.0005 & 0.0000 && -0.0008 & 0.0000 && -0.0010 & 0.0000\\
				$\tau_{32}$ && -1.5409 & 0.1675 && -0.3375 & 0.0312 && -0.4012 & 0.0406 && -0.3987 & 0.0298\\
				\hline
				AIC && 3756.218 &&& \textbf{3538.333} &&& 3600.380 &&& 3658.380 \\
				BIC && 4422.084 &&& \textbf{4218.071} &&& 4280.118 &&& 4351.990 \\
				RI  && 0.5123   &&&  0.5214  &&&  \textbf{0.5323}  &&& 0.5118 \\ 
				JCI && 0.3676   &&&  0.3847  &&&  0.4029  &&& 0.3713 \\	 
				\hline
			\end{tabular}   
	}}
\end{table}
\noindent
Based on the AIC and BIC criteria, the TP-SMN-MoECPLE model demonstrates a superior fit to the data compared to competing models. The results indicate that education has a significant positive influence on the wages of husbands. This effect is most pronounced in the first expert, followed by the second, and finally the third, in a decreasing order of magnitude. Furthermore, the first expert exhibits pronounced asymmetry ($\hat \gamma_1=0.27$) and heavy-tailed behavior ($\hat \nu_1=2.05$). In contrast, the second and third experts display weak asymmetry ($\hat \gamma_2=0.41$ and $\hat \gamma_3=0.62$) but remain relatively heavy-tailed ($\hat \nu_2=7$ and $\hat \nu_3=13$) (see also right part of Figure \ref{fig7}).

\noindent
The left part of Figure \ref{fig7} illustrates the non-parametric density estimates of the fitted values for each of the three experts. The distinct separation and minimal overlap between these density curves provide clear visual evidence of a mixture structure comprising three distinct experts. Moreover, the shapes of these densities effectively capture the underlying probabilistic characteristics of each component, visually confirming the presence of skewness and heavy-tailed behaviour within the distributions of the individual experts.

\begin{figure}[htbp]
	\centering
	\includegraphics[height=7cm, width=14cm]{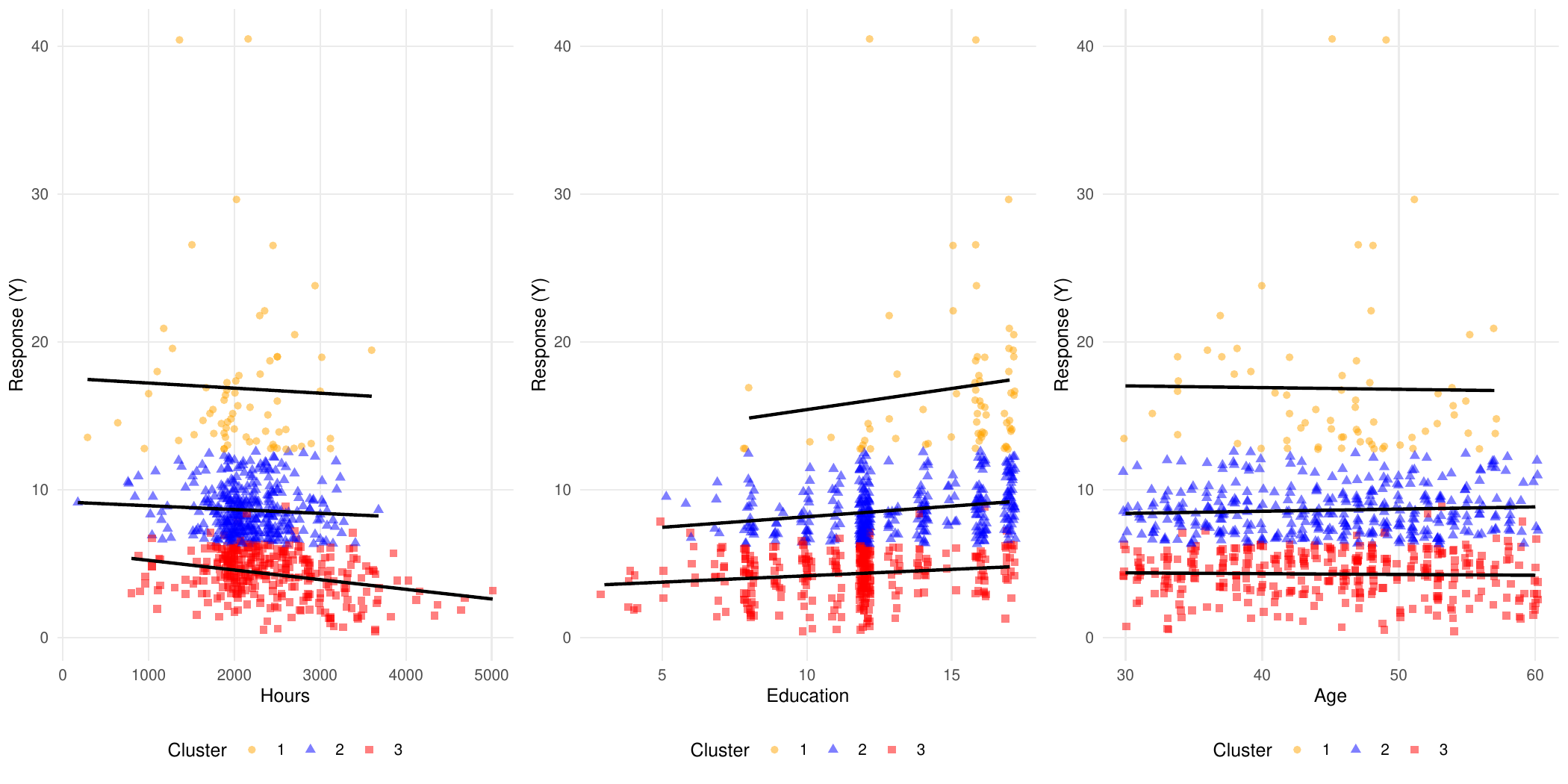} 
	\vspace{0.2cm} 
	\caption{Pairwise clustering of the husband's wage rate data based on the fitted TP-SMN-MoCPLE model.}
	\label{fig8}
\end{figure}
\noindent
Figure \ref{fig8} displays the scatter plots of the response variable against the covariates---hours, education, and age---where the observations are categorized into three distinct clusters based on the fitted model. To visualize the underlying relationship within each group, linear regression lines are superimposed on the plots. These regression lines demonstrate appropriate trends within each cluster, showing consistency with the estimated experts. The clear separation of the clusters and the distinct trajectories of the fitted lines across the different covariates provide compelling visual evidence that the proposed three-component mixture model effectively captures the heterogeneity of the data. This structure confirms that the clustering derived from the model is well-defined and meaningful.

\section{Conclusions and discussions}\label{sec6}
This paper introduces a novel robust class of Mixture of Partially Linear Experts (MoCPLE) models tailored for censored data, grounded in the family of two-piece scale-mixture of normal distributions. The proposed TP-SMN-MoCPLE framework extends the classical Mixture of Experts (MoE) paradigm, effectively addressing complex data challenges such as non-linearity, asymmetry, heavy-tailed disturbances, outliers, and censoring mechanisms. This model is highly general and encompasses various existing frameworks as special cases, including symmetric and asymmetric classical MoE models, finite mixtures of Partially Linear Models (PLM), Mixture Regression Models (MRM), and models designed for censored data. Furthermore, the incorporation of covariate-dependent gating functions facilitates the automatic selection of experts, thereby enhancing classification performance. By leveraging the TP--SMN family, the proposed model accommodates both light- and heavy-tailed distributions as well as skewness; notably, the zero-skewness constraint naturally recovers the well-known standard SMN family. A key advantage of this class is the availability of closed-form expressions for conditional expectations, which significantly streamlines the definition and estimation of the model. Notably, prior to this work, censored models based on skewed distributions possessing closed-form conditional expectations and allowing for exact Maximum Likelihood Estimators (MLE) without resorting to Monte Carlo approximations were non-existent. This contrasts with the present study, which offers this distinct computational advantage. The proposed methodology has been implemented in R software, and the corresponding computational code is available from the authors upon request.

\bibliographystyle{elsarticle-harv}
\bibliography{Reff}

\appendix
\def\theequation{A.\arabic{equation}}
\setcounter{equation}{0}

\section{Conditional expectations of the TP-SMN distributions}\label{appa}
{\bf Uncensored cases:}
\medskip\\
For the uncensored data $y$, we have $\rho =0$. Therefore, we have the following results:
\begin{lemma}
	\label{lem:A1}
	Let $Y \sim \mathcal{TP-SMN}(\mu,\sigma,\gamma,\nu)$, then considering the stochastic representation \eqref{eq:6}, conditional expectation $\hat{uf} = E\left[UF^{-2}| Y=y\right]$ for the TP-SMN distribution members, which are needed the uncensored part of the data on the EM-algorithm, are given by:
	\begin{itemize}
		\item[$\bullet$] TP-N: $\hat{uf} = \mathbb{I}_{(-\infty, \mu]} (y)$,
		\item[$\bullet$] TP-T: $\hat{uf} = \sum_{f} \frac{\nu+1}{f^2 (\nu+d_f^2)} \mathbb{I}(F=f)$,
		\item[$\bullet$] TP-SL: $\hat{uf} = \sum_{f} \frac{2\nu+1}{f^2 d_f^2} \frac{P_1 (\nu+3/2, d_f^2/2)}{P_1 (\nu+1/2, d_f^2/2)} \mathbb{I}(F=f)$,
		\item[$\bullet$] TP-CN: $\hat{uf} = \sum_{f} \frac{\nu\tau^2 e^{-(\tau d_f^2)/2} + (1-\nu)e^{-(d_f^2)/2}}{\tau\nu e^{-(\tau d_f^2)/2} + (1-\nu)e^{-(d_f^2)/2}} \frac{1}{f^2} \mathbb{I}(F=f)$,
	\end{itemize}
	where $d_f^2 = \left(\frac{y-\mu}{\sigma f}\right)^2$ and $P_x(a,b)$ denote the cdf of the Gamma$(a,b)$ distribution evaluated at $x$. Also considering $S = \mathrm{Sign}(F)$, we have $\hat s= E\left[S| Y=y\right]=\mathbb{I}(Y\ge \mu)-\mathbb{I}(Y< \mu)$.

\end{lemma}

{\bf Censored cases:}
\medskip\\ 
In the censored cases, we have $\rho =1$. First, we define the expectation functions 
\[
E_\phi(m,s) = E\left[U^m \phi\big(s\sqrt{U}\big) \right]\quad \text{and} \quad 
E_\Phi(m,s) = E\left[U^m \Phi\big(s\sqrt{U}\big) \right],
\]
which are needed at follow, and for each specail cases of the TP--SMN class of distributions are given by: 
\begin{itemize}
	\item[$\bullet$] \textbf{TP-N:}
	\begin{align*}
		E_\phi(m,s) = \phi(s),\\ E_\Phi(r,h)=\Phi(s),
	\end{align*}
	
	\item[$\bullet$] \textbf{TP-T:}	
	\begin{align*}
		&E_{\phi}(m,s) = \frac{\Gamma\left(\frac{\nu+2m}{2}\right)}{\sqrt{2\pi}\, \Gamma\left(\frac{\nu}{2}\right)} \left(\frac{\nu}{2}\right)^{\frac{\nu}{2}} \left(\frac{s^2+\nu}{2}\right)^{-\frac{\nu+2m}{2}},\\
		&E_{\Phi}(m,s) = \frac{\Gamma\left(\frac{\nu+2m}{2}\right)}{\Gamma\left(\frac{\nu}{2}\right)} \left(\frac{\nu}{2}\right)^{-m} F_{P VII}(s; \nu+2m, \nu),
	\end{align*}
	where $F_{PVII}(\cdot;\nu,\delta)$ denotes the cdf of Pearson type $VII$ distribution,
	\item[$\bullet$] \textbf{TP-SL:}
	\begin{align*}
		&E_{\phi}(m,s) = \frac{\nu}{\sqrt{2\pi}} \left(\frac{s^2}{2}\right)^{-(\nu+m)} \Gamma\left(\nu+m, \frac{s^2}{2}\right),\\
		&E_{\Phi}(m,s) = \frac{\nu}{\nu+m} F_{SL}(s; \nu+m),
	\end{align*}
	where $\Gamma(a,b) = \int_{0}^{\infty} e^{-t} t^{a-1} dt$ is the incomplete gamma function and $F_{SL}(\cdot)$ is the cdf of the slash distribution,
	\item[$\bullet$] \textbf{TP-CN:}
	\begin{align*}
		&E_{\phi}(m,s) = \nu \tau^m \phi(s\sqrt{\tau}) + (1-\nu)\phi(s),\\
		&E_{\Phi}(m,s) = \nu \tau^m \Phi(s\sqrt{\tau}) + (1-\nu)\Phi(s).
	\end{align*}	
\end{itemize}

\begin{lemma}
	\label{lem:A2}
	Let $Y \sim \mathcal{TP-SMN}(\mu,\sigma,\gamma,\nu)$, 
	$\mathcal{A} = (a < \mu \le b)$, $f_1 = -(1-\gamma)$, $f_2 = \gamma$, $a_1^* = \frac{a-\mu}{\sigma \left| f_1 \right|}$, $b_1^* = 0$, $a_2^* = 0$, $b_2^* = \frac{b-\mu}{\sigma \left| f_2 \right|}$, $\eta(\mathcal{A}) = F(b; \mu, \sigma, \gamma, \nu) - F(a; \mu, \sigma, \gamma, \nu)
	$, where $F$ is the cdf of the TP-SMN family in \eqref{eq:5}, and
	\begin{align*}
		&\zeta_1(a,b) = E_{\phi}\left(\frac{1}{2}, a\right) - E_{\phi}\left(\frac{1}{2}, b\right),\\
		&\zeta_2(a,b) = E_{\Phi}(0, b) - E_{\Phi}(0, a) + a E_{\phi}\left(\frac{1}{2}, a\right) - b E_{\phi}\left(\frac{1}{2}, b\right).
	\end{align*}
	Then we have:
	\begin{align*}
		&E_{F,Y,U} \left[ F^{-2} U \mid Y \in \mathcal{A} \right] = \frac{1}{\eta(\mathcal{A})} \left[ \frac{\zeta_0(a_1^*, b_1^*)}{f_1^2} + \frac{\zeta_0(a_2^*, b_2^*)}{f_2^2} \right],\\
		&E_{F,Y,U} \left[ F^{-2} U Y \mid Y \in \mathcal{A} \right] = \frac{1}{\eta(\mathcal{A})} \left\{ \mu \left[ \frac{\zeta_0(a_1^*, b_1^*)}{f_1^2} + \frac{\zeta_0(a_2^*, b_2^*)}{f_2^2} \right] + \sigma \left[ \frac{\zeta_1(a_1^*, b_1^*)}{|f_1|} + \frac{\zeta_1(a_2^*, b_2^*)}{|f_2|} \right] \right\},\\
		&E_{F,Y,U} \left[ F^{-2} U Y^2 \mid Y \in \mathcal{A} \right] = \frac{1}{\eta(\mathcal{A})} \bigg\{ \mu^2 \left[ \frac{\zeta_0(a_1^*, b_1^*)}{f_1^2} + \frac{\zeta_0(a_2^*, b_2^*)}{f_2^2} \right] + 2\mu\sigma \left[ \frac{\zeta_1(a_1^*, b_1^*)}{|f_1|} + \frac{\zeta_1(a_2^*, b_2^*)}{|f_2|} \right] \nonumber \\
		&\qquad + \sigma^2 \left[ \zeta_2(a_1^*, b_1^*) + \zeta_2(a_2^*, b_2^*) \right] \bigg\}.
	\end{align*}
	Also we have $\hat s= E\left[S| Y \in \mathcal{A}\right]=
	\frac{\eta(\mu,b)-\eta(a,\mu)}{\eta(a,b)}
	\mathbb{I}(a<\mu<b)+
	\mathbb{I}(\mu \le a)-\mathbb{I}(b\le \mu)$. 
	
\end{lemma}

\begin{lemma}
	\label{lem:A3}
	Considering the assumptions of Lemma \ref{lem:A2}, when  
	$\mu \notin \mathcal{A} = (a, b)$, then for
	
	\textbf{Case 1:} $b \le \mu$, $F = -(1-\gamma)$ with probability $1$.
	In the above formulas we have:
	\[
	\mathcal{A}^* = \left( a^* = \frac{a-\mu}{\sigma(1-\gamma)}, \quad b^* = \frac{b-\mu}{\sigma(1-\gamma)} \right),
	\]
	
	\textbf{Case 2:} $\mu < a$, $F = \gamma$ with probability $1$.
	We have:
	\[
	\mathcal{A}^* = \left( a^* = \frac{a-\mu}{\sigma\gamma}, \quad b^* = \frac{b-\mu}{\sigma\gamma} \right).
	\]
	then for both cases, we have:
	\begin{align*}
		&E_{F,Y,U} \left[ F^{-2} U \mid Y \in \mathcal{A} \right] = \frac{1}{\eta(\mathcal{A})} \left[ \frac{\zeta_0(a^*, b^*)}{F^2} \right],\\
		&E_{F,Y,U} \left[ F^{-2} U Y \mid Y \in \mathcal{A} \right] = \frac{1}{\eta(\mathcal{A})} \left\{ \mu \left[ \frac{\zeta_0(a^*, b^*)}{F^2} +  \right] + \sigma \left[ \frac{\zeta_1(a^*, b^*)}{|F|} \right] \right\},\\
		&E_{F,Y,U} \left[ F^{-2} U Y^2 \mid Y \in \mathcal{A} \right] = \frac{1}{\eta(\mathcal{A})} \left\{ \mu^2 \left[ \frac{\zeta_0(a^*, b^*)}{F^2} \right] + 2\mu\sigma \left[ \frac{\zeta_1(a^*, b^*)}{|F|} \right] \nonumber + \sigma^2 \left[ \zeta_2(a^*, b^*) \right] \right\}.
	\end{align*}
	Also for case 1, we have $\hat s= E\left[S| Y \in \mathcal{A}\right]=1$ and for case 2, $\hat s= E\left[S| Y \in \mathcal{A}\right]=-1$. 
\end{lemma}

\end{document}